*universe* MDPI

*Article*

# Multiverse Predictions for Habitability: Number of Potentially Habitable Planets

**McCullen Sandora** [1,2]

[1] Institute of Cosmology, Department of Physics and Astronomy, Tufts University, Medford, MA 02155, USA; mccullen.sandora@gmail.com

[2] Center for Particle Cosmology, Department of Physics and Astronomy, University of Pennsylvania, Philadelphia, PA 19104, USA

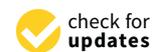

**Abstract:** How good is our universe at making habitable planets? The answer to this depends on which factors are important for life: Does a planet need to be Earth mass? Does it need to be inside the temperate zone? are systems with hot Jupiters habitable? Here, we adopt different stances on the importance of each of these criteria to determine their effects on the probabilities of measuring the observed values of several physical constants. We find that the presence of planets is a generic feature throughout the multiverse, and for the most part conditioning on their particular properties does not alter our conclusions much. We find conflict with multiverse expectations if planetary size is important and it is found to be uncorrelated with stellar mass, or the mass distribution is too steep. The existence of a temperate circumstellar zone places tight lower bounds on the fine structure constant and electron to proton mass ratio.

## 1. Introduction

This paper is a continuation of [1], which aims to use our current understanding from a variety of disciplines to estimate the number of observers in a universe $N_{\text{obs}}$, and track how this depends on the most important microphysical quantities such as the fine structure constant $\alpha = e^2/(4\pi)$, the ratio of the electron to proton mass $\beta = m_e/m_p$, and the ratio of the proton mass to the Planck mass $\gamma = m_p/M_{pl}$. Determining these dependences as accurately as possible allows us to compare the measured values of these constants with the multiverse expectation that we are typical observers within the ensemble of allowable universes [2]. In doing so, there remain key uncertainties that reflect our ignorance of what precise conditions must be met in order for intelligent life to arise. Rather than treating this obstacle as a reason to delay this endeavor until we have reached a more mature understanding of all the complex processes involved, we instead view this as a golden opportunity: since the assumptions we make alter how habitability depends on parameters, sometimes drastically, several of the leading schools of thought for what is required for life are incompatible with the multiverse hypothesis. Generically, if we find that our universe is no good at a particular thing, then it should not be necessary for life because, if it were, we would most likely have been born in a universe which is better at that thing. Conversely, if our universe is preternaturally good at something, we expect it to play a role in the development of complex intelligent life, otherwise there would be no reason we would be in this universe. The requirements for habitability are in the process of being determined with much greater rigor through advances in astronomy, exoplanet research, climate modeling, and solar system exploration, so we expect that in the not too distant future our understanding of the requirements for intelligent life will be much more complete. At this stage of affairs, then, we are able to use the multiverse hypothesis to generate predictions for which of these habitability criteria will end up being true. These will either be vindicated, lending credence to the multiverse hypothesis, or not, thereby falsifying it.

In this work, we define the habitability of a universe $\mathbb{H}$ as the total number of observers it produces. In estimating the number of observers the universe contains, a great many factors must be taken into consideration. Thankfully, there has long been a useful way of organizing these factors: the Drake equation, which in a slightly modified form along the lines of [3] reads

$$\mathbb{H} = \int_{\lambda_{\min}}^{\infty} d\lambda \ p_{\text{IMF}}(\lambda) \times N_\star \times f_{\text{p}}(\lambda) \times n_{\text{e}}(\lambda) \times f_{\text{bio}}(\lambda) \times f_{\text{int}}(\lambda) \times N_{\text{obs}}(\lambda). \quad (1)$$

Here, $N_\star$ is the number of stars in the universe, $f_{\text{p}}$ is the fraction of stars containing planets, $n_{\text{e}}$ is the average number of habitable planets around planet-bearing stars, $f_{\text{bio}}$ is the fraction of planets that develop life, $f_{\text{int}}$ is the fraction of life bearing worlds that develop intelligence, and $N_{\text{obs}}$ is the number of intelligent observers per civilization. We have included dependence of the size of the host star $\lambda = M_\star/((8\pi)^{3/2} M_{pl}^3/m_p^2)$, in order to more accurately reflect the fact that these quantities may depend on this. We then integrate over the stellar initial mass function given in [4], which approximates a broken power law with turnover at $0.2 M_\odot$. This can be related to the probability of being in our universe by incorporating the relative occurrence rates of different universes as $P \propto p_{\text{prior}} \mathbb{H}$. It was argued in [1] that a reasonable choice of prior is given by $p_{\text{prior}} \propto 1/(\beta\gamma)$. However, this is ultimately set by physics at high energies, and so may in principle be something else.

Previously, the fact that the strength of gravity $\gamma$ can be two orders of magnitude higher caused the biggest problems with finding a successful criterion, since most stars are in universes with stronger gravity. Though we had set out to focus solely on the properties of stars, it was only when we weighted the habitability of a system by the total entropy processed by its planets over its entire lifetime that we hit upon a fully satisfactory criterion. This was also reliant on the condition that starlight be in the photosynthetic range, colloquially referred to as 'yellow' light here: conservatively, this corresponds to the 600–750 nm range. Relaxing this range to be from 400–1100 nm will not qualitatively affect our results. The other factors we considered may be freely included at will without hindering this conclusion. For the majority of this paper, we take the entropy and yellow light conditions as our baseline minimal working model, and incorporate factors that influence the availability and properties of planets to determine how these alter our estimates for habitability. While our previous analysis was not heavily reliant on cutting edge results from the field of astronomy, our understanding of planets, from their population statistics to their formation pathways, has undergone rapid expansion in the past decade, and a state-of-the-art analysis needs to reflect that.

To this end, we begin by estimating the fraction of stars with planets in Section 2. Most notable is the recent determination of a threshold metallicity below which rocky planets are not found [5], as well as the understanding of the origin of this threshold. We also find the conditions for the lifetime of massive stars to be shorter than the star formation time and the fraction of galaxies able to retain metals to be sizable, but find that these only impose mild constraints. Using these allows us to determine the dependence of $f_{\text{p}}$ on the underlying physical parameters. Additionally, we incorporate the fraction of systems that host hot Jupiters into our analysis, and find conditions for this process to not wreck all planetary systems.

In Section 3, we turn to the average number of habitable planets $n_{\text{e}}$. Two commonly used requirements for a planet to be habitable are that it needs to be both terrestrial and temperate, and so we optionally include both of these when estimating habitability. Recent results indicate that the distribution of rocky planets peaks at $1.3 R_\oplus$ [6–9], which is rather close to the terrestrial radius capable of supporting an Earthlike atmosphere. We track how this quantity changes in alternate universes, and what implications this effect has on the number of habitable planets in these universes. Additionally, we track the location and width of the temperate zone and compare this to the typical inter-planet spacing that results from the dynamical evolution of stellar systems, which provides a rough estimate for the probability that a planet will end up in the temperate zone. Finally, we discuss the importance of planet migration and how this changes in other universes.

An appendix is provided to collect the relevant formulas for the dependence on the physical constants on the variety of processes and quantities that are needed, in order to avoid distraction from the main text.

The overarching message we derive from this analysis is that the presence of planets is not that important in determining our location in this universe, a direct consequence of the fact that the presence of planets is a nearly universal phenomenon throughout the multiverse. Including these effects barely alters the probabilities we derived before. We find that most effects act as thresholds, serving to limit the allowed parameter range rather than alter the probability distribution of observing any particular value. We find several new anthropic bounds, including the most stringent lower bound on the electron to proton mass ratio in the literature. We find that these results are relatively insensitive to the exact models of planet formation and occurrence rate used, and so are robust to these current uncertainties.

This is not the first work to address the question of whether planets are still present for alternative values of the physical parameters. Limitations on the strength of gravity and electromagnetism imposed by the existence of habitable planets was investigated in [10], where it was found that long lived, temperate, terrestrial planets can exist over a wide range of parameter space. This was continued in [11] with the investigation of the influence of the density of galaxies on planetary stability, again finding a broad allowable parameter region. Our current work is novel in not just examining the possibility of the existence of planets with desirable properties, but also taking care to incorporate modern theories of planet formation into determining whether planets with these properties are indeed produced.

Taken together, we analyze 12 distinct possible criteria for habitability in this paper (not counting migration or the different views in the planetary size distribution we consider). Coupled to the 40 we considered in [1], this represents a total of 480 different hypotheses to compare. The quantities used to compute these are displayed in Table 1 for convenience.

**Table 1.** The quantities computed in this work. Here, $Q$ is the amplitude of perturbations, $\kappa$ parameterizes the density of galaxies, $\lambda$ parameterizes stellar size, GI stands for giant impact formation mechanism, and iso stands for isolation production mechanism.

| Quantity | Description | Expression |
|---|---|---|
| $f_{\text{gal}}$ | fraction of stars in galaxies that retain supernova ejecta | $\text{erfc}(4.1 Q^{-1} \alpha^2 \beta^{5/3})$ |
| $f_{\text{2nd gen}}$ | fraction of stars born after supernova enrichment | $\exp(-0.24 \kappa^{3/2} \alpha^{-7/4} \beta^{-1/8} \gamma^{-1})$ |
| $f_Z$ | fraction of stars with high enough $Z$ for planets | $\theta(1 - 0.038 \lambda^{3/4} \alpha^{-3} \beta^{-1/2} \gamma^{1/2})$ |
| $f_{\text{hj}}$ | fraction of stars without hot Jupiters | $1 - 2.3 \times 10^8 \kappa^2 \lambda^2 \alpha^{-13/8} \beta^{-3}$ |
| $n_p$ | average number of planets around a star | GI: Equation (25), iso: Equation (31) |
| $f_{\text{terr}}$ | fraction of terrestrial planets | GI: Equation (27), iso: Equation (30) |
| $f_{\text{temp}}$ | fraction of temperate planets | $0.0053 \kappa^{-1/2} \lambda^{-1.77} \alpha^{11/2} \beta^{7/4} \gamma^{-5/8}$ |

## 2. Fraction of Stars with Planets $f_P$

Two of the factors in the Drake equation regard the existence of planets, which will be considered in turn in this paper. The first quantity to determine is $f_P$, the fraction of stars that form planets. Here, we represent this as a product of factors:

$$f_P = f_{\text{gal}} \times f_{\text{2nd gen}} \times f_Z \times (f_{\text{hj}})^{p_{\text{hj}}}. \tag{2}$$

In succession, we have: $f_{gal}$, the fraction of stars in galaxies large enough to retain supernova ejecta, $f_{\text{2nd gen}}$, the fraction of stars born after supernova enrichment, $f_Z$, the fraction of stars born with high enough metallicity for planets to form, and an optional $f_{hj}$, the fraction of stars that do not produce hot Jupiters. Here, the exponent $p_{hj} \in \{0, 1\}$ is introduced as a choice of whether to include this last criterion or not. The other two are not treated as optional.

*2.1. What Sets the Size of the Smallest Metal-Retaining Galaxy?*

The requirement for a galaxy to be habitable is that it must retain its supernova ejecta in order to reprocess it into another round of metal-rich stars [12]. This sets a minimum galactic mass by the condition that the velocity of supernova ejecta is less than the escape velocity,

$$v_{SN}^2 \sim \frac{G\,M_{\text{ret}}}{R_{\text{ret}}}. \tag{3}$$

This is the asymptotic speed the supernova ejecta attains, and, to find this, a bit of the ejecta dynamics must be used. The initial speed can be found from energy balance [13]: this can be written in the form

$$v_{SN}^0 \sim \sqrt{\frac{T_{SN}}{A\,m_p}}, \tag{4}$$

where the temperature of the supernova is set by the Gamow energy, which is the amount required to overcome the repulsive nuclear barrier and force fusion, $T_{SN} \sim \alpha^2 m_p$. The mass of a typical particle in the ejecta is related to the atomic number $A \sim 50$, which cannot conceivably vary by much. We find that $v_{SN}^0 \sim 0.03$. However, as the ejecta moves through the intergalactic medium, it cools and slows until it merges completely. The asymptotic speed is that at which the temperature becomes equal to (about an order of magnitude less than) the hydrogen binding energy $T_{H_2} \sim \alpha^2 m_e/32 \sim 10^4$ K, below which Hydrogen becomes predominantly neutral and no more cooling takes place [14].

How far does the ejecta of a supernova spread before it completely merges with the interstellar medium, and, more importantly for our purposes, why? The observed value is around 100 pc [15], and this is after the blast has gone through several successive phases. The first is known as the blast wave phase, where the ejecta spread out at their initial velocity for roughly 100 yr, traveling a total of a few pc. After the amount of interstellar material encountered rivals the initial mass of the ejecta, which occurs at $d_{ST} \sim (M_{ej}/\rho_{gal})^{1/3}$, the blast enters the self-similar Sedov–Taylor phase, where the blast slows and expands considerably. According to standard theory [15], the self-similarity of the dynamics dictates that the temperature of the blast falls off as $T \propto d^{-3}$. During this phase, the velocity will decrease with distance as $v(d) = 2/5(d_{ST}/d)^{5/2}$ until the snowplow phase begins, at which point the speed essentially does not decrease any further. The snowplow phase occurs after the temperature reaches the molecular cooling threshold, where it expands by a factor of a few until the density of the material falls to that of the surrounding medium, at which point it is completely merged. Since the bulk of the expansion takes place in the Sedov–Taylor phase, the size of the blast will be dictated by the dynamics that take place there, and so the ultimate speed is given by

$$v_{SN} = v_{SN}^0 \left(\frac{T_{SN}}{T_{H_2}}\right)^{5/6}. \tag{5}$$

Using the above relations, the asymptotic speed of supernova ejecta is found to be

$$v_{SN} = 2.4\,\alpha\,\beta^{5/6}. \tag{6}$$

Now, we can find an expression for the minimum mass by using $M_{\min} \sim \rho_{\text{gal}} R^3$. Using the density of galaxies given in the appendix, this gives

$$M_{\text{ret}} \sim \frac{M_{pl}^3 v_{\text{SN}}^3}{\rho_{\text{gal}}^{1/2}} = 90.1 \frac{\alpha^3 m_e^{5/2} M_{pl}^3}{\kappa^{3/2} m_p^{9/2}}, \quad (7)$$

where the coefficient in the last expression has been chosen to reproduce the observed minimal mass of $M_{\text{ret}} = 10^{9.5} M_\odot$ [16]. As explained in the appendix, the quantity $\kappa$ determines the density of galaxies in terms of cosmological parameters.

While this critical mass is important conceptually, it is not as relevant to the retention of ejecta as the gravitational potential itself, which directly sets the escape velocity of the overdensity. The fraction of initial overdensities that exceed a potential of a given strength $\Phi$ is given by the Press–Schechter formalism as $f = \text{erfc}(\Phi/(\sqrt{2}Q))$ [17], where $Q$ is the primordial amplitude of perturbations. Usually, this expression is immediately expressed in terms of mass and density, but this more primitive form will suffice for our purposes. With this, we can derive the fraction of matter that resides in potential wells deep enough to produce a second generation of stars as

$$f_{\text{gal}} = \text{erfc}\left(4.1 \frac{\alpha^2 \beta^{5/3}}{Q}\right). \quad (8)$$

Perhaps somewhat interestingly, this does not depend on the strength of gravity $\gamma$ at all. This effectively acts as a step function, severely diminishing the habitability of universes where $M_{\text{ret}} > M_{\text{typ}}$. Note that this is a pessimistic estimate, as it ignores the potential for subsequent evolution that causes potential wells to deepen with time. Nevertheless, it is a very mild bound, and so a more thorough treatment is not called for.

*2.2. Is Massive Star Lifetime Always Shorter Than Star Formation Time?*

Since the formation of metal-rich systems is reliant on the evolution of the first stars through to their completion, if the lifetime of massive stars exceeds the duration of star formation, then no systems will form with any substantial metallicity. The second generation stars are not necessarily enriched enough to produce planets, but this serves as a sufficient condition for planets to be formed at all. Then, the fraction of stars with planets can be estimated as those that are born after a few massive stellar lifetimes have elapsed. It is worth considering how these two timescales compare for general values of the physical constants.

The star formation rate, averaged throughout the universe, is found to decline exponentially, as gas is depleted from the initial reservoir [18]. The most naive treatment one can perform is to relate the timescale of this depletion to the free fall time of a galaxy, $t_{\text{dep}} = 1/(\epsilon_{\text{SFR}}\sqrt{G\rho})$, where for simplicity the efficiency coefficient is taken to not vary with parameters. Then, the fraction of stars born after a time $t$ is $f_\star(t) = e^{-t/t_{\text{dep}}}$.

This needs to be compared to the lifetime of massive stars, where massive here is taken to mean large enough to become a type II supernova. This threshold is eight solar masses in our universe, and is set by the inner core having a high enough temperature to undergo carbon fusion [19]. As such, this threshold is parametrically similar to the minimum stellar mass, which was shown in [20] to scale as $\lambda \propto \alpha^{3/2} \beta^{-3/4}$, which equated the Gamow energy of fusion with the internal temperature of the star. The resultant mass is about two orders of magnitude larger than the minimal mass, stemming from the larger repulsive barrier for large nuclei. Then, using the stellar lifetime from [1], we find

$$\frac{t_{\text{SN}}}{t_{\text{dep}}} = 0.24 \frac{\kappa^{3/2}}{\alpha^{7/4} \beta^{1/8} \gamma}. \quad (9)$$

The normalization has been set to the value of 0.01. This has been a bit loose on several counts, namely the assumption that the second stars are always metallic enough to form planets, and the neglect of even higher mass stars, which have correspondingly shorter lifetimes. However, in practice, these worries are of no consequence, precisely because the scales are separated by such a large amount. We find that, for all practical values of the parameters, the star formation timescale exceeds the lifetime of massive stars, so that $f_{\text{2nd gen}} \sim f_\star(t_{\text{SN}})$ is always close to 1.

*2.3. What Is the Metallicity Needed to Form Planets?*

Terrestrial planets must be formed out of heavy elements, and though even a minuscule amount would suffice in terms of actual planetary mass, the planet formation processes within the protoplanetary accretion disk can only occur after a high enough metallicity is reached. A recent analysis [5] has done a nice job characterizing the metallicity needed (as a function of stellar mass, and distance to the host star) in order for planets to form out of the initial protoplanetary disk.

The underlying physical picture is that there are two timescales, the lifetime of the disk $t_{\text{disk}}$ and the dust settling time $t_{\text{dust}}$. If the second exceeds the first, the disk will dissipate before sufficient dust may settle into the midplane, and will disperse without forming planets. Both of these depend on metallicity monotonically, and so only above some critical value will the conditions for planetary formation hold. We detail the scaling of each in turn.

The physics of accretion disks was first laid out in [21]. An initially uncollapsed cloud first condenses, and then through angular momentum transfer begins to form a disk. The disk remains around the star until either UV light from the star photoevaporates the gas or it escapes thermally, and the disk evaporates. This phase of evolution occurs on the viscous timescale of the disk, which is much smaller than the total disk lifetime [22].

Though the precise time at which this crossover occurs depends on the exact mechanism of photoevaporation (X-ray versus extreme ultraviolet, other stellar sources in clusters, presence of high activity tau phase) the drop-off of accretion is set by the disk's secular evolution time [23], with crossover occurring on the order of this time. Therefore, the only relevant physics that needs to be kept track of is the free fall time. Since this is given by a cloud that has approximately virialized after Jeans collapse, it is simply set by the condition that the gravitational energy is equal to the thermal energy. Then, the accretion rate is given by $\dot{M} = c_s^3/G \sim 10^{-6} M_\odot/\text{yr}$, and the timescale is given by $t_{\text{disk}} = M_\star/\dot{M} \sim \text{Myr}$. The sound speed and the temperature are given by bremsstrahlung from molecular line cooling, as found in the appendix. Then,

$$t_{\text{disk}} = 5.5 \times 10^5 \frac{\lambda \, m_p^{1/4} \, M_{pl}}{\alpha^3 \, m_e^{9/4}} \left(\frac{Z}{Z_\odot}\right)^p. \tag{10}$$

In [24], the metallicity dependence was studied, where it was found that cooling quickens with metallicity $Z$. The exact process was open to interpretation, with a handful of viable candidate mechanisms, but the overarching explanation is that the less shielding there is, the faster the disk will dissipate. Observationally and theoretically, it was found that the scaling with metallicity is consistent with $p = 1/2$, and so we will adopt this for our analysis.

The dust settling timescale is dictated by the rate at which dust grains sink into the midplane from their initial positions in the protoplanetary disk. It is given in terms of the Keplerian timescale of the disk, $t_{\text{dust}} = 0.72 t_{\text{Kepler}}/Z$, as derived in [5]. There it was related to the growth rate of grains, and found to depend inversely on the metallicity, since only dust can participate in the accretion process. This is a function of orbital distance, but specifying to planets that form within what will become a temperate orbit for simplicity, we have

$$t_{\text{dust}} = 3.0 \times 10^5 \frac{\lambda^{17/8} \, m_p^{7/4} \, M_{pl}^{1/4}}{\alpha^{15/2} \, m_e^3} \frac{Z_\odot}{Z}. \tag{11}$$

Here, we are explicitly assuming that significant migration does not occur, which would alter the metallicity needed for planet formation, given its dependence on orbit.

Equating these two timescales yields the critical metallicity

$$Z_{\min} = 4.2 \times 10^{-4} \lambda^{3/4} \frac{\gamma^{1/2}}{\alpha^3 \beta^{1/2}}. \tag{12}$$

The critical value is found to be $Z_{\min} = 6.3 \times 10^{-4}$. As compared to the solar value $Z_\odot = 0.02$, we find $Z_{\min} = 0.03 Z_\odot$ [5].

This may be used to determine the fraction of stars that host planets by considering the amount of stars that are formed above this metallicity. This requires a model for how metallicity builds up inside a galaxy (of a given mass, as well as the distribution of galaxy masses). A full calculation of this sort would take us too far afield here so we will return to it in a later publication. However, a very reasonable approximation is to compare this threshold metallicity to the asymptotic metallicity a galaxy attains—this usually affords percent-level accuracy. In our universe, this is found to be $Z_\infty = 0.011$ [16,25], which is actually a factor of two below solar due to the intrinsic scatter in metallicities. This asymptotic value is set by the fraction of stellar mass that is transformed into heavy elements during stellar fusion, and is not expected to depend sensitively on the fundamental constants over the ranges considered.

If attention is restricted to solar mass stars, the requirement that the threshold metallicity be lower than the asymptotic value equates to $\alpha^3 \beta^{1/2} \gamma^{-1/2} > 0.80$, which is most sensitive to the fine structure constant. Leaving the other parameters fixed, this gives $\alpha > 1/361$, which is stronger than the bound $\alpha > 1/685$ based on galactic cooling found in [26,27]. A more forgiving bound is found using the smallest stellar mass, which is $\alpha^{5/2} \beta^{17/12} \gamma^{-2/3} > 0.32$. This latter boundary is displayed in Figure 1, where the distribution of observers is plotted as a function of the three variables $\alpha$, $\beta$ and $\gamma$.

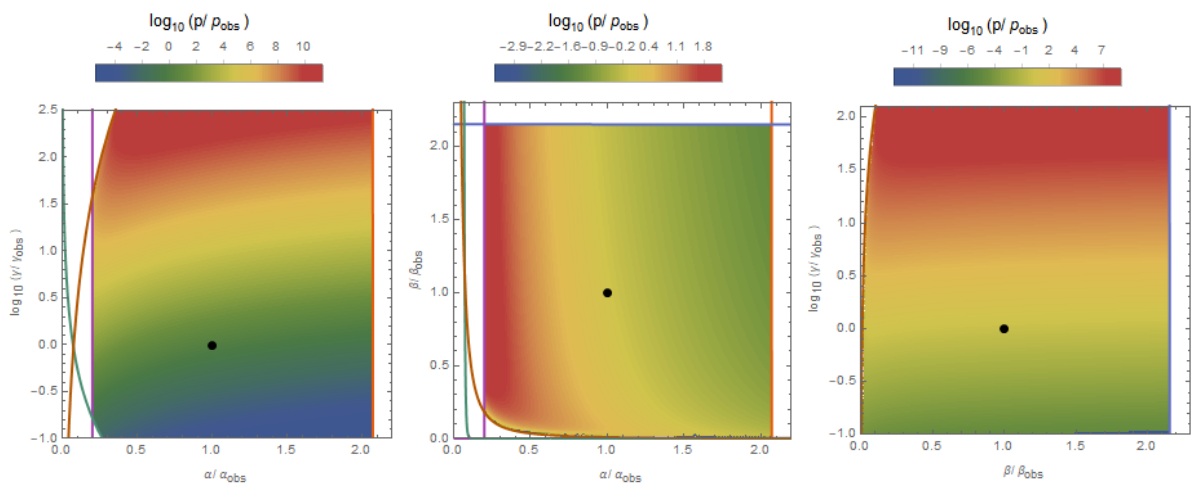

**Figure 1.** The distribution of observers when taking metallicity effects into account, in the $\alpha$–$\gamma$, $\alpha$–$\beta$, and $\beta$–$\gamma$ subplanes. What is plotted is the logarithm of the probability of measuring any value of the coupling constants, with red being more probable than blue, and the black dot corresponding to our observed values. Included thresholds are an upper bound on $\alpha$ for hydrogen stability, as well as a lower bound for galactic cooling, and an upper bound on $\beta$ for proton–proton fusion. The teal and brown curves correspond to the metallicity and supernova lifetime thresholds, respectively, and the supernova retention threshold is not relevant in this range.

Incorporating the above effects determining the fraction of stars that host planets, and again using the entropy and yellow light conditions, we can find that the probability of observing our particular values of each constant. These are defined as $\mathbb{P}(x_{\rm obs}) = \min\{P(x > x_{\rm obs}), P(x < x_{\rm obs})\}$ for any observable quantity $x$, the others being integrated over, and the probability of measuring any value given by Equation (1). With this, we have

$$\mathbb{P}(\alpha_{\text{obs}}) = 0.19, \quad \mathbb{P}(\beta_{\text{obs}}) = 0.45, \quad \mathbb{P}(\gamma_{\text{obs}}) = 0.32. \tag{13}$$

These values are indistinguishable from those that were found in [1] without including these effects. We conclude that the presence of planets is a generic feature throughout most of the multiverse, and it does not alter where we expect to be situated in the slightest[1].

*2.4. Are Hot Jupiter Systems Habitable?*

The constraints above were all quite mild, indicating that the presence of planets is a fairly generic feature of the multiverse. However, we can make a further refinement by incorporating one of the most famous statistical correlations in the field of exoplanets, the hot Jupiter–metallicity correlation [28]. This finds that the fraction of stars that possess hot Jupiters, that is, Jupiter sized planets on orbits extremely close to the star increase with metallicity as $Z^2$. The general (though not universal— [29]) consensus is that these planets must have formed in the outer system before moving inward, to avoid the necessity of a disk that would be so massive as to be unstable [30]. In this scenario, the migration of the planet through the inner solar system would have certainly ejected any preexisting planets from their orbits (or worse), precluding them from sustaining life. However, the authors in [31] hypothesize that this entire process could happen early enough that the main stage of planet formation could occur after this migration had already taken place. If this turns out to be the case, then there may in fact be no correlation between hot Jupiters and habitability (barring other factors that may impact habitability [32]). If one wishes to include this effect, however, then the fraction of stars that host Earths and not Hot Jupiters is given by

$$f_{\text{hj}} = 1 - \frac{\bar{Z}^2}{Z_{\text{max}}^2}. \tag{14}$$

Here, we have used the mean metallicity rather than averaging this fraction over the metallicity distribution, but this approximation is sufficient for a first analysis. The normalization $Z_{\text{max}}$ is the threshold above which the stellar system is almost assured to possess a hot Jupiter. It is set to reproduce the observed abundance of hot Jupiter systems of 3% as $Z_{\text{max}} = 5.77 Z_\odot$. In a multiverse setting, we would expect to inhabit a universe where $Z_{\text{max}}$ is safely above the average metallicity, beyond which any further increase would result in little increase in habitability. Somewhat in line with this expectation, then, is the fact that in our universe hot Jupiters exist in only a few percent of systems, and mainly in those that are highly metal-rich.

To investigate whether this is the result of some selection effect, we must know what determines this metallicity. The functional dependence is a clue: as the effect becomes more pronounced with the square of the nongaseous material present, this is indicative of an interaction process. What remains, however, is the question of whether this migration is a result of planet–planet or planet–disk interactions. In fact, both explanations have been considered in the literature: references may be found in [33] [2]. The planet–planet hypothesis is supported by the fact that the eccentricities of observed hot Jupiters are correlated with metallicity as well, indicating a more chaotic, violent origin, rather than the steady, deterministic process indicative of planet–disk interaction. Additionally, [35] note a substantial misalignment between the orbits of known hot Jupiters and the spins of their host stars, which is most easily explained through a violent migration scenario. However, systems like WASP-47 [36], which possess both a hot Jupiter and smaller companions, demonstrate that more dynamically quiet migration pathways are possible, if not necessarily the norm. Here, we only expound upon the planet–planet scenario, though the others could just as readily be incorporated into our analysis.

---

[1] The code to compute all probabilities discussed in the text is made available at https://github.com/mccsandora/Multiverse-Habitability-Handler.

[2] A third explanation was additionally given in [34] that the disk dispersal timescale increases with metallicity, allowing a longer period of accretion onto seed cores.

We start by determining the value of $Z_{max}$ due to planet–planet interactions, as proposed in [37]. As noted in [38], the coexistence of hot Jupiters and low mass planets is impossible in this paradigm, as migration occurs after the disk has dissipated. From [33], this is set by the expected number of Jupiter mass planets initially formed in the outer system. They provide a framework for estimating this by determining the probability that a core will attain Jupiter mass as a result of planetesimal accretion as $p \propto \Delta M/M_{crit}$, where $M_{crit} \sim 10 M_\oplus$ is the mass above which runaway gas accretion is possible and $\Delta M$ is the typical total accreted mass. For this, we use the analytic expression for the accretion rate from [39],

$$\dot{M} \approx (18\pi)^{1/3} \frac{G\, M_{crit}\, M_\star^{1/3}\, \Sigma_{\text{planetesimals}}}{\rho^{1/3}\, a}. \tag{15}$$

This employs the strong gravitational focusing limit, and treats the typical relative velocities of planetesimals as roughly given by the Hill velocity. This can be used to determine the total mass accreted by simply multiplying by the disk lifetime given in Equation (10) (making use of the simplifications that the nonlinear oligarchic regime is not quite reached, and the initial isolation timescale is small compared to the disk lifetime). The probability that there will be at least two gas giants to trigger the instability will scale as $p(\geq 2) \sim N_{jup}^2 p^2$, where $N_{jup}$ is the typical number of planets in the outer system, and we have assumed that $p$ is small. The quantity $N$ can be found by dividing the total mass of the planetary disk by the typical mass of a planet at the typical location of formation. Here, we use the initial seed being set by the isolation mass, have fixed the orbital radius to be given by the snow line, and have taken the disk temperature to be given by viscous accretion, all of which are discussed in the appendix. This gives

$$N_{jup} \sim \frac{M_{disk}}{M_{crit}} \sim 2.2 \times 10^{-5} \frac{\lambda}{\alpha^{3/2} \beta^{3/4}}. \tag{16}$$

This scales linearly with stellar mass, in agreement with the observations in [40]. The maximal value of metallicity is found to be

$$Z_{max} \sim \frac{1}{N_{jup}} \frac{M_{crit}}{\dot{M}\, t_{disk}} = 2.3 \times 10^{-8} \frac{\alpha^{13/6} \beta^{3/2}}{\kappa\, \lambda}. \tag{17}$$

This quantity is somewhat sensitive to both $\alpha$ and $\beta$, but not at all to $\gamma$. This also defines a stellar mass $\lambda_{hj} = 2.3 \times 10^{-6} \alpha^{13/6} \beta^{3/2}/\kappa$: stars above this mass, equal to 11 $M_\odot$ for our values, will always host hot Jupiters. This will be below the smallest stellar mass if $789 \alpha^{8/27} \beta < 1$, which will occur when the electron to proton mass ratio is about 10 times smaller. However, this criteria does not alter the probabilities much:

$$\mathbb{P}(\alpha_{obs}) = 0.18, \quad \mathbb{P}(\beta_{obs}) = 0.44, \quad \mathbb{P}(\gamma_{obs}) = 0.31. \tag{18}$$

With this, notice that the probabilities are only changed from Equation (13) by a few percent. Thus, even when including the demand that hot Jupiter systems are uninhabitable, the fraction of systems with planets seems to be relatively insensitive to the physical constants.

We now turn to the next factor in the Drake equation, which deals with the characteristics of planets, rather than just their presence.

## 3. Number of Habitable Planets per Star $n_e$

Next, we focus on the number of habitable planets per star, $n_e$. The determination of habitability may depend on many factors, such as amount of water, eccentricity, presence of any moons, magnetic field, distance from its star, atmosphere, composition, etc. Here, we focus on two: temperature and size, and determine the fraction of stars that have planets with each of these characteristics.

As usual, it is possible that habitability is completely independent from these properties; which viewpoint one adopts depends on how habitable one expects environments without liquid surface water and thin atmospheres can be. In this work, we remain agnostic to either expectation and report the number of observers for all combinations of choices, where a planet will only be habitable if it is approximately Earthlike, and where the size and/or temperature of the planet have no effect on its habitability.

It should be noted that we are restricting our attention here to surface dwelling life on planets orbiting their star. Thus, life in subsurface oceans and/or on icy moons like Enceladus [41], or even more exotic types of life (e.g., [42,43]), are not considered. It is our plan to consider these alternative environments in future work.

The number of habitable planets can be broken down as

$$n_\text{e} = n_\text{p} \times (f_\text{terr})^{p_\text{terr}} \times (f_\text{temp})^{p_\text{temp}}. \tag{19}$$

Here, the average total number of planets around a star is $n_\text{p}$. The fraction of terrestrial mass planets is denoted $f_\text{terr}$, and $f_\text{temp}$ is the fraction of planets that reside within the temperate zone. The exponents $p_i \in \{0,1\}$ parameterize the choice of whether to include these conditions in the definition of habitability or not. These quantities all depend on stellar mass, giving preference to large stars because they make larger planets, and small stars in that their temperate zone is wider compared to the interplanetary spacing.

Estimating these quantities is somewhat muddled by the current uncertainties in planet formation theory. Not only is the distribution of planet masses contested in the literature, but the exact formation pathways, as well as the physics that dictates the results, is not completely settled. Where we come across disagreement, we separately try each proposal, in order to understand the sensitivity of our analysis to present uncertainties. While the results for the overall probabilities can vary by a factor of 2, the upshot is that our estimates are relatively robustx.

*3.1. Why Does Our Universe Naturally Make Terrestrial Planets?*

The size of a planet is of crucial importance because it dictates what kind of atmosphere it can retain. If it is too small, all atmospheric gases will eventually escape, whereas if it is too large, it will retain a thick hydrogen and helium envelope, leading to a runaway growth process. Terrestrial planets must have a very specific size in order that the escape velocity exceeds the thermal velocity for heavy gases such as $H_2O$, $CO_2$ and $N_2$, but not that of the lightest gas H and He. In our universe, and for temperatures within the range where liquid surface water is possible, this restricts the range of planetary radii to be within 0.7 and 1.6 that of Earth's [44]. This is a narrow sliver compared to the eight orders of magnitude mass range of spherical, non-fusing bodies, ranging from the potato radius of 200 km to 10 Jupiter masses. In terms of fundamental parameters, this requirement gives the mass to be

$$M_\text{terr} = 92 \, \frac{\alpha^{3/2} M_{pl}^3 \, m_e^{3/4}}{m_p^{11/4}}. \tag{20}$$

The coefficient has been set to reproduce Earth's mass, but the allowed spread in masses is taken to be between 0.3–4 $M_\oplus$.

There are compelling arguments that complex may only be possible on terrestrial planets, with atmospheres composed of only heavy gases [45]. Any smaller, and the planet would be Marslike, an apparently barren wasteland incapable of sustaining any appreciable liquid ar atmosphere. The other extreme would be Neptunelike, with its hellish surface temperatures, pressures and wind speeds. Of course, these arguments may be misguided, but here we explore the consequences of adopting them for the multiverse computations we perform.

It is important to note that the conditions that set the presence of atmospheres are completely separate from the physics that dictates the size of planets, which is set by the clumping of the initial circumstellar disk[3]. Nonetheless, the observed population of rocky planets is thought to peak at only slightly super-Earth mass, making the production of terrestrial planets the norm for stars throughout the universe. To be fair, the current exoplanet samples are biased towards large mass planets and become very incomplete below Earth mass [48], but a number of different groups have concluded that a detectable turnover is present near Earth masses: the authors of Ref. [9] find a good fit to a log-normal distribution that peaks at $1.3R_\oplus$. In Ref. [7] a Rayleigh distribution with width $3M_\oplus$ is used. The authors of Ref. [8] advocate for a broken power law with turnover at $5M_\oplus$. It was noted in Ref. [6] that the distribution appears to be flat below $2.8R_\oplus$. Use of these differing proposed distributions make very little difference to our final outcome.

However, not everyone is convinced that the mass distribution exhibits a peak, and even if there is one, it is just as reasonable to assume that there are many more smaller mass planets for every planet of Earth size, as the plethora of small asteroids and comets in our system indicates. Because of the incompleteness of current exoplanet surveys for small mass planets, there is room for such disagreement at the current moment. Additionally, even if the mass peak is real, it is only observed for close in exoplanets, and so requires an extrapolation to Earthlike orbits, where different dynamics may be at play. One possibility is that the peak at super-Earth mass may be due to their enhanced migration capability [49]. We will consider each scenario in turn.

3.1.1. What Sets the Size of Planets?

Why is the turnover so nearly equal to Earth mass planets, out of the potentially eight orders of magnitude that could have been selected instead? Simulations provide a means to address this question: it was found in Ref. [50] that the mass of planets is directly proportional to the amount of initial material present in the disk, so that increasing disk mass makes larger, rather than more, planets. In this scenario, nearly all the material initially present in the disk eventually gets constituted into planets, with negligible (perhaps a factor of two, but not an order of magnitude) losses throughout the evolution of the system. Determining the final planet mass in this setup requires knowledge of the initial disk mass, as well as the fraction of material within the inner solar system. This boundary is set by the snow line, the difference in composition interior and exterior to which dictates the formation of rocky versus icy planets. In the following, we adopt the conventional view that little migration takes place during planet formation, and comment further on alternative pathways when migration is discussed.

Current observations indicate that the initial mass of heavy elements in protoplanetary disks is roughly proportional to disk mass, $M_{\rm disk} = 0.01 M_\star$ [51]. In Ref. [52], a stronger dependence was found, but the scatter of 0.5 dex is larger than the trend, and the observed trend was suggested to possibly be due to a selection effect arising from processing into undetectably large grains, so we omit this stronger scaling for now. For solar mass stars, this works out to be roughly $10 M_{\rm Jupiter}$, or $3300 M_\oplus$. This mass is distributed out to a radius of $\sim$100 AU, which is set by the conservation of angular momentum, and the initial size of the collapsing cloud. From the appendix, we arrive at the following expression for disk size:

$$r_{\rm disk} = 3.6 \times 10^{-6} \frac{\lambda^{1/3}}{\kappa} \frac{M_{pl}}{m_p^2}. \tag{21}$$

---

[3]   This is not entirely true: there is known to be some feedback between the irradiation of initial atmospheres from the host star that favors both small and large atmospheres [46]. While this leads to the interesting bimodal distribution of observed radii [47], this is driven by atmospheric size, and is certainly not enough to affect the terrestrial core.

Universe **2019**, *5*, 157　　　　　　　　　　　　　　　　　　　　　　　　　　　　　　　　　　　　　　　　　12 of 34This may be significantly altered if the young star is in a dense environment [53], but this scaling will suffice for our purposes.

To determine the amount of material present within the snow line, one must take the surface density profile into account. For this, we use the expression for $\Sigma(a)$ found in the appendix, which is inversely proportional to $a$. Using these, the typical mass of an interior planet becomes

$$M_{\text{inner}} \sim \eta_< \frac{a_{\text{snow}}}{R_{\text{disk}}} M_{\text{disk}}, \tag{22}$$

where we have included the quantity $\eta_< \sim 0.25$ to account for the difference in composition interior to the snow line [54].

The location of the snow line is the point in the disk beyond which water condenses, equal to 2.7 AU in our solar system [55]. In order to determine its location, the temperature as a function of radius must be used, which will depend on the dominant source of heating. For most of the present paper, we use the temperature given by viscous accretion, and find

$$a_{\text{snow}} = 30.4 \frac{\lambda \, M_{pl}}{\alpha^{5/3} \, m_e^{5/4} \, m_p^{3/4}}. \tag{23}$$

Note that, perhaps unsurprisingly, the snow line is situated outside the temperate zone for all relevant parameter values. The dependence on stellar mass was found taking $\dot{M} \propto \lambda^2$, but is generically found to scale as $a_{\text{snow}} \propto \lambda^{2/3-2}$ [56]. This is now enough to determine the parameter dependence of $M_{\text{inner}}$, which will ultimately be used to derive the expected number of planets per star as a function of these parameters.

Though we tend to favor accretion dominated disks throughout this work, irradiation from the central star can actually play a significant role as well [22]. If this is the dominant mode of heating, then the snow line will instead be given by the expression $a_{\text{snow}} = 297 \lambda^{43/30} \alpha^{-4} m_e^{-4/3} m_p^{-1/3} M_{pl}^{2/3}$. Though this is functionally quite similar to the accretion dominated case from above, in Table 2, we investigate the effect of assuming this form instead. Actually, both are almost equally relevant in determining the position of the snow line, which helps to greatly complicate the disk structure (as well as enhance the variability between different star systems [57]). Additionally, irradiation from other neighboring stars may be important as well, especially in clusters [58], but we do not consider this contribution in this work.

**Table 2.** Display of the insensitivity of the probabilities of our observables to the choices made regarding planet formation. In addition to the options displayed, they may have been combined, but this would belabor the point.

| Choices | $\mathbb{P}(\alpha_{\text{obs}})$ | $\mathbb{P}(\beta_{\text{obs}})$ | $\mathbb{P}(\gamma_{\text{obs}})$ |
|---|---|---|---|
| standard | 0.229 | 0.260 | 0.409 |
| Rayleigh distribution | 0.229 | 0.251 | 0.411 |
| irradiation | 0.255 | 0.320 | 0.426 |
| with no $\lambda$ dependence | 0.380 | 0.254 | 0.007 |
| shot noise | 0.232 | 0.260 | 0.306 |

Determining the average planet mass is somewhat involved, given the many distinct stages of growth that occur as microscopic dust grains agglomerate to the size of planets. For reviews of this multi-stage process, see [59–61]. In brief, planetesimals form characteristic masses as the final outcome of pebble (or dust) accretion. This arrangement is unstable, and ultimately leads to a phase of giant impacts, wherein planetesimals collide together to form full planets.

An estimate for the maximum mass a planet can attain after a phase of growth through chaotic giant impacts was found in [62]. There, they assumed no migration, small eccentricity, and determined the width of the 'feeding zone' to be $\Delta a = 2v_{\text{esc}}/\Omega$ by noting that within this region planet–planet interactions result in collisions rather than velocity exchange. This yields

$$M_{\text{planet}} \sim \left( \frac{4\pi \Sigma a^{5/2} \rho^{1/6}}{M_\star^{1/2}} \right)^{3/2}, \qquad (24)$$

where $\rho$ is the average density of the planet and $\Sigma$ is the disk density. With this, we can use the appendix to reinstate parameter dependence into the expressions for the average number of planets (normalized to 3 for the solar system) as well as the typical planet mass:

$$n_{\text{p}} = 0.0061 \frac{\alpha^{4/3} \beta^{13/16}}{\kappa^{1/2} \lambda^{5/6}}, \qquad (25)$$

$$\frac{M_{\text{planet}}}{M_{\text{terr}}} = 1.6 \times 10^6 \frac{\kappa^{3/2} \lambda^{5/2}}{\alpha^{9/2} \beta^{45/16}}. \qquad (26)$$

The latter has quite a steep dependence on stellar size. This is expected from the simulations [50], with a dependence closer to linear, but is not particularly observed in exoplanet catalogs due to large scatter [63]. In our calculations, we explore the effect of ignoring this dependence altogether, displayed in Table 2. A full treatment would take the dependence on semimajor axis into account, rather than simply evaluating at the snow line: in fact, this would be somewhat unnecessary, as the scatter observed in exoplanet surveys, simulations, and indeed within the solar system masks any dependence that may exist.

For the fraction of planets that are terrestrial, we use the log-normal distribution of [9] since this is what is expected of the core accretion process, though we check that the actual distribution used does not alter the outcome much. Under these assumptions, we can find the probability that a planet will be terrestrial as

$$f_{\text{terr}} = \frac{1}{2}\text{erf}\left( \frac{\log\left(\frac{4 M_{\text{terr}}}{M_{\text{planet}}}\right) + \frac{1}{2}\sigma_M^2}{\sqrt{2}\,\sigma_M} \right) - \frac{1}{2}\text{erf}\left( \frac{\log\left(\frac{.3 M_{\text{terr}}}{M_{\text{planet}}}\right) + \frac{1}{2}\sigma_M^2}{\sqrt{2}\,\sigma_M} \right). \qquad (27)$$

This function peaks at $M_{\text{planet}} \sim M_{\text{terr}}$, and approaches are when $M_{\text{planet}}$ is very different from $M_{\text{terr}}$. Being a two-parameter distribution, this requires not just the mean, but also the variance. As it is not currently known what sets this quantity, here we explore two options: the first uses maximization of entropy production to set $\sigma_M = 1/\sqrt{6}$, which is fairly widely observed in natural processes [64]. This estimate should occur for large systems, but for small systems one would expect the variance to be set by shot noise instead, $\sigma \sim \sqrt{M_{\text{iso}}/M_{\text{inner}}}$, making use of the isolation mass defined in the next section. The dependence on the various parameters is displayed in Figure 2.

The probabilities of observing the observed values of our constants are computed for the various choices we made in Table 2. These can be compared with Equation (13) that only considered the fraction of stars with planets. The most significant change for our most favored prescription is the probability of observing our electron to proton mass ratio, which is decreased by less than a factor of 2. Such insensitivity hardly constitutes any evidence for whether life should only appear on terrestrial planets. More interestingly, if the dependence on stellar mass is neglected, our strength of gravity becomes quite uncommon to observe. This gives us strong reason to suppose that, if life requires terrestrial planets, we will begin to see a correlation between the two soon, and, if we don't, then life requiring Earth mass planets is incompatible with the multiverse at the $2.7\sigma$ level.

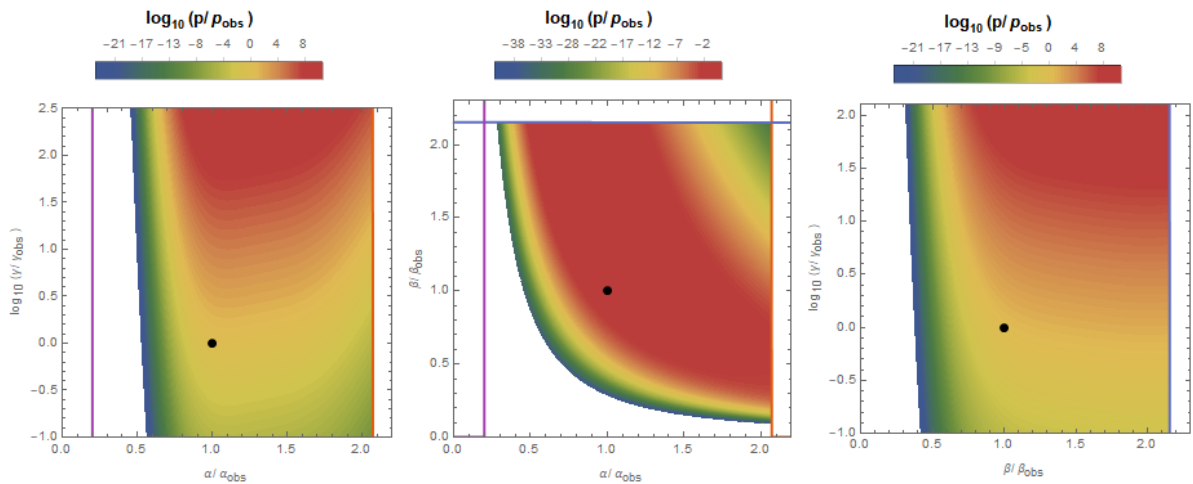

**Figure 2.** The distribution of observers if life can only arise on a terrestrial planet in the $\alpha$–$\gamma$, $\alpha$–$\beta$, and $\beta$–$\gamma$ subplanes, size being dictated by a giant impact phase. Here, we have used a log-normal distribution with $\sigma = 1/\sqrt{6}$, excluded $\lambda$ dependence on the average planet mass, and assumed accretion dominated disks.

Aside from this, though, the choices we made to come up with these estimates do not affect the outcome very much at all. On the one hand, this is disappointing, as the stronger the dependence these probabilities have on the assumptions of planet formation, the stronger our predictions can be about which to expect to be dominant. However, this is also heartening: because the current uncertainties about planet formation do not affect the outcome all that much, we are able to trust the broad conclusions we have reached a bit better.

3.1.2. Is Life Possible on Planetesimals?

The mass discussed above really refers to the maximal planet mass of a system, which form as a result of the secondary stage of collisions after the isolated planetesimals form. However, this agglomeration will likely not completely deplete the system of its primordial planetesimals, and so there are also expected to be numerous smaller planets accompanying each large one, as is the case in and around the solar system's asteroid belt. If these smaller bodies are considered as potential abodes for life as well, the distribution continues past Earth masses, rather than having a peak there.

As a planet is condensing out of the protoplanetary disk, it eventually reaches what is termed as the pebble isolation mass, which from the appendix is given by

$$M_{\text{iso}} = 2.0 \times 10^8 \, \frac{\kappa^{3/2} \lambda^2 M_{pl}^3}{\alpha^{5/2} m_e^{15/8} m_p^{1/8}}. \tag{28}$$

The isolation mass is a function of the semi-major axis, but if we evaluate it at the edge of the inner system given by Equation (23), we find, using the value for the disk surface density from the Appendix A,

$$\frac{M_{\text{iso}}}{M_{\text{terr}}} = 2.2 \times 10^6 \, \frac{\kappa^{3/2} \lambda^2}{\alpha^4 \beta^{21/8}}. \tag{29}$$

Here, the density of the galaxy comes into play in setting the outer edge of the disk. The dependence on stellar mass in this expression is quite close to that found in [55], $M_{\text{iso}} \propto \lambda^{7/4}$.

It remains to specify the distribution of planetary masses in order to find the fraction that are terrestrial in this picture. It is generically expected to take a power law form that continues to the small mass cutoff, so that $N(M) = (M_{\text{iso}}/M)^q$. However, different authors prefer different values for the slope: Ref. [65] find $q = 0.31 \pm 0.2$. Ref. [66] find a nearly scale invariant distribution for the radius, which translates into $q = 0.30 \pm 0.03$ if we use $M \propto r^3$. Refs. [39,67] find $q = 0.6$ for simulations of planetesimals, and Ref. [54] extrapolate from the known asteroid population to find $q = 1$. Ref. [68] favors a value of $q = 2$ from population synthesis methods. Here, we report with various values of $q$ to investigate its influence on the probabilities. The fraction of terrestrial planets is then

$$f_{\text{terr}} = \min\left\{1, \left(\frac{M_{\text{iso}}}{0.3\, M_{\text{terr}}}\right)^q\right\} - \min\left\{1, \left(\frac{M_{\text{iso}}}{4\, M_{\text{terr}}}\right)^q\right\}. \tag{30}$$

This is an increasing function of stellar mass until $M_{\text{iso}} > 0.3 M_{\text{terr}}$, reflecting the expectation [69] that earthlike planets should be rare among low mass stars. With this prescription, stars above a certain mass will not produce any earthlike planets because the isolation mass will exceed the largest terrestrial planet size. This defines the largest viable stellar mass $\lambda_{\text{iso}} = 0.0013\kappa^{-3/4}\alpha^2\beta^{21/16}$, which corresponds to $6.3\, M_\odot$ in our universe. This will only exceed the minimal stellar mass if $\alpha^{1/2}\beta^{33/16}\kappa^{-3/4} > 169$. This condition is most sensitive to the electron to proton mass ratio and, holding the other constants fixed, will be violated if it drops to about 11% of its observed value.

For $q < 1$, the average mass for this distribution is formally infinite, which presents a problem for using our expression for the expected number of planets in a system as given by $n_p = M_{\text{inner}}/\langle M_p \rangle$. However, the total mass in the inner disk introduces a large mass cutoff, for which we have

$$n_p = \frac{q-1}{q} \frac{\left(\frac{M_{\text{inner}}}{M_{\text{iso}}}\right) - \left(\frac{M_{\text{inner}}}{M_{\text{iso}}}\right)^{1-q}}{1 - \left(\frac{M_{\text{inner}}}{M_{\text{iso}}}\right)^{1-q}}. \tag{31}$$

This interpolates between $(q-1)/q M_{\text{inner}}/M_{\text{iso}}$ for $q > 1$ and $(q-1)/q$ for $q < 0$. The ratio of masses is

$$\frac{M_{\text{inner}}}{M_{\text{iso}}} = 0.0017 \frac{\alpha^{5/6}\beta^{5/8}}{\kappa^{1/2}\lambda^{1/3}}, \tag{32}$$

thus that this equates to 30 with our constants. With this viewpoint, the distribution of observers is plotted in Figure 3.

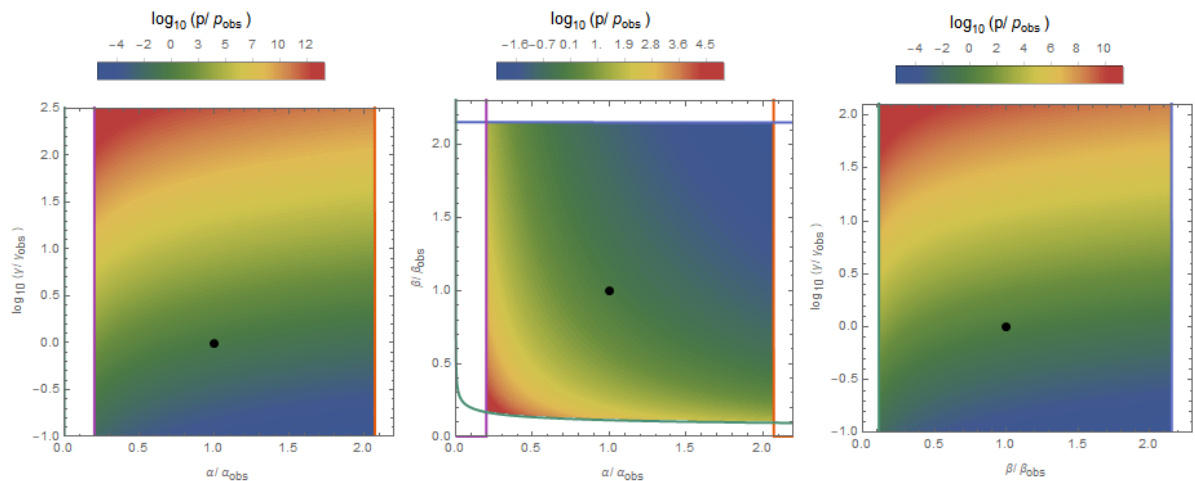

**Figure 3.** The distribution of terrestrial planetesimals that result from isolated accretion in the $\alpha$–$\gamma$, $\alpha$–$\beta$, and $\beta$–$\gamma$ subplanes, without the subsequent phase of giant impacts. The slope here is $q = 1/3$. The teal line represents the region where the largest star capable of hosting terrestrial planets is smaller than the smallest possible star.

The overall probabilities are calculated for different representative values of the power law in Table 3. It can be seen that, for increasing $q$, the probability for $\alpha$ increases, while the other two decrease.

In particular, for $q = 2$, the probability of observing our value of the electron to proton mass ratio is disfavored by $2\sigma$.

Table 3. Display of the probabilities of our observables for different values of the power law slope.

| Exponent | $\mathbb{P}(\alpha_{\text{obs}})$ | $\mathbb{P}(\beta_{\text{obs}})$ | $\mathbb{P}(\gamma_{\text{obs}})$ |
|---|---|---|---|
| $q = 1/3$ | 0.234 | 0.419 | 0.436 |
| $q = 2/3$ | 0.302 | 0.260 | 0.424 |
| $q = 1$ | 0.362 | 0.157 | 0.333 |
| $q = 2$ | 0.469 | 0.044 | 0.228 |

*3.2. Why Is the Interplanet Spacing Equal to the Width of the Temperate Zone?*

Perhaps the most commonly employed habitability criteria is that a planet must be positioned a suitable distance away from its host star to maintain liquid water on its surface. This assumption is so pervasive that this region is usually referred to as the circumstellar habitable zone. In the spirit of remaining agnostic toward the conditions required for life, we will adhere to the recently proposed renaming as the 'temperate zone' [70]. If this is indeed essential for life, then the expected number of habitable planets orbiting a star will depend both on the interplanetary spacing, as well as the width and location of the temperate zone. A rather clement feature of our universe is that the width of the temperate zone is comparable to the interplanetary spacing (for sunlike stars). Because of this, it is relatively common that one of the planets in any stellar system is situated inside the temperate zone, no matter its particular arrangement. This could be contrasted to the hypothetical case where the temperate zone were much narrower than the interplanetary spacing, in which case the odds of a planet being situated inside it would be quite low. However, this coincidence of distance scales is not automatic: both these quantities are dependent on the underlying physical parameters, and so in universes with different parameter values the expected number of potentially habitable planets per star will be altered. We go through these length scales in turn, and then fold them into our estimate for the overall habitability of the universe.

The boundaries of the temperate zone depend on the characteristics of the planet in question, such as the atmospheric mass and composition [71], its orbital period [72], etc. However, these details only alter the location of the temperate zone to subleading order, and do not affect the scaling with fundamental parameters we are interested in. If the planet is assumed to be a simple blackbody, then the temperature is set solely by the amount of incident flux. In this case, the location of the temperate zone will be

$$a_{\text{temp}} = \frac{1}{2} \frac{T_\star^2}{E_{H_2O}^2} R_\star = 7.6 \frac{\lambda^{7/4} m_p^{1/2} M_{pl}^{1/2}}{\alpha^5 m_e^2}. \tag{33}$$

Albedo and greenhouse effects will change the coefficient, but not the overall scaling. Determining the width of the temperate zone entails finding the temperatures at which runaway climate processes occur, but we will now argue that both the inner and outer edge are dictated by (broadly speaking) the same underlying physical process of phase change, and so the width of the temperate zone will scale in the same way as its location.

The inner edge of the temperate zone is set by the runaway greenhouse effect: this occurs when a temperature threshold is crossed that allows an appreciable amount of water vapor to be sustained in the atmosphere. Since this serves to trap infrared light from escaping to space, this will increase the temperature further, in turn driving a further increase in atmospheric water vapor. Once the atmosphere is comprised primarily of water, it will be photodissociated and/or escape to space, leaving the Earth in a dry and Venus-like state [73]. Since there is always some level of outward flux, the ocean will escape into space eventually if a long enough time has elapsed. Therefore, the exact threshold for this process is defined as when the timescale for this process is of the order of one billion years, which in turn will depend on the mass of the planet's ocean. However, key to our discussion is that the change in the atmospheric water fraction occurs very abruptly when the temperature crosses the latent heat of vaporization, going from equilibrium values of $10^{-5} - 1$ in the span of 250–420 K. Because of this, the exact mass of the planet, ocean or atmosphere will only play a subleading role in determining this threshold, which instead is dictated solely by molecular processes. Since this transition is set by what is essentially the intermolecular binding energy, this scales identically with the condition for liquid water. For Earth, this occurs at a temperature of 330 K, which corresponds to 0.95 AU [74].

Similarly, the outer edge of the temperate zone is set by the runaway icehouse process. The temperature of a planet is set not only by the incident flux, but also by the carbon dioxide content content of the atmosphere and albedo. Carbonate weathering is an important regulatory feedback mechanism that serves to stabilize the temperature of a planet to remain within the temperate range by adjusting atmospheric carbon dioxide content to compensate for a change in stellar flux [75] (it does this because liquid water is necessary for efficient weathering to occur). However, this only works to an extent, since beyond a critical concentration atmospheric $CO_2$ increases the albedo of a planet, leading to a runaway icehouse effect [73]. Because the amount of atmospheric carbon dioxide is also set by reactions due to intermolecular forces, it scales the same as above. Therefore, both the inner and outer edges of the temperate zone are set by molecular binding energies, and so the width will always be of the same order as the mean[4].

The exact delineations are subject to the uncertainties of the atmospheric model used, but, for definiteness, we take $\Delta R_{\rm HZ} = 0.73$ AU, from [71].

The next task is to determine how far apart planets typically reside from each other. It was hypothesized in [77] that stellar systems are dynamically packed, in that they are filled to capacity, and the insertion of any additional planet would render the system unstable. Though this may not strictly hold all the time [78], this spacing is roughly observed in our solar system [79], in Kepler data on multiplanetary systems [80], and found to occur naturally in simulations in [81]. The essential idea is that planetary scattering either ejects or collides excess planets from an initially overpacked system until the remainder are far enough apart to be stable on the timescale of the system. The stability condition is that planets are further than a certain multiple of Hill radii away, nominally around 10, and it was found in [80] that the distribution of separations was a shifted Rayleigh distribution with usual separation $21.7 \pm 9.5 R_{\rm Hill}$.

As mentioned in [62], the typical multiple of mutual Hill radii is not universal, but can be shown to depend on the width of a planet's 'feeding zone' to be given by $(\rho a^3/M_\star)^{1/4}$, where $\rho$ is the density of matter and $a$ is the semimajor axis. Though above we were interested in the typical size of a planet during this process, here we may use this condition to find the typical spacing as

$$a_{\rm spacing} \sim \left(\frac{16\pi \Sigma a^{11/2} \rho^{1/2}}{M_\star^{3/2}}\right)^{1/2} = 1430 \frac{\kappa^{1/2} \lambda^{169/48} \gamma^{1/8}}{\alpha^{21/2} \beta^{15/4}} \quad (34)$$

This was evaluated at the center of the temperate zone to give the ratio of these two length scales in terms of fundamental constants. Since this ratio loosely sets the fraction of planets within the temperate zone, we arrive at

$$f_{\rm temp} \sim \frac{\Delta a_{\rm temp}}{a_{\rm spacing}} = 0.0053 \frac{\alpha^{11/2} \beta^{7/4}}{\kappa^{1/2} \lambda^{85/48} \gamma^{5/8}}. \quad (35)$$

---

4   This neglects other potential thresholds, such as the inner boundary set by the photosynthetic threshold of carbon dioxide abundance, which is remarkably close to the inner boundary discussed here [76].

Set this to the value of 1.6 in our solar system. Notice that, for $\alpha$ or $\beta$ much smaller or $\gamma$ much larger, the fraction of temperate planets will be diminished. Additionally, this quantity is larger for smaller mass stars, reflecting the comparatively broad temperate zones there.

This distribution is illustrated in Figure 4. The corresponding probabilities for observing our values of the constants are

$$\mathbb{P}(\alpha_{obs}) = 0.24, \quad \mathbb{P}(\beta_{obs}) = 0.37, \quad \mathbb{P}(\gamma_{obs}) = 0.15. \tag{36}$$

Though the probabilities shift around by a factor of a few, the inclusion of this criterion does not actually affect the results very much. This is a generic conclusion, even when this is included with other criteria. Whether or not this condition is essential for life, the multiverse hypothesis is relatively insensitive to it.

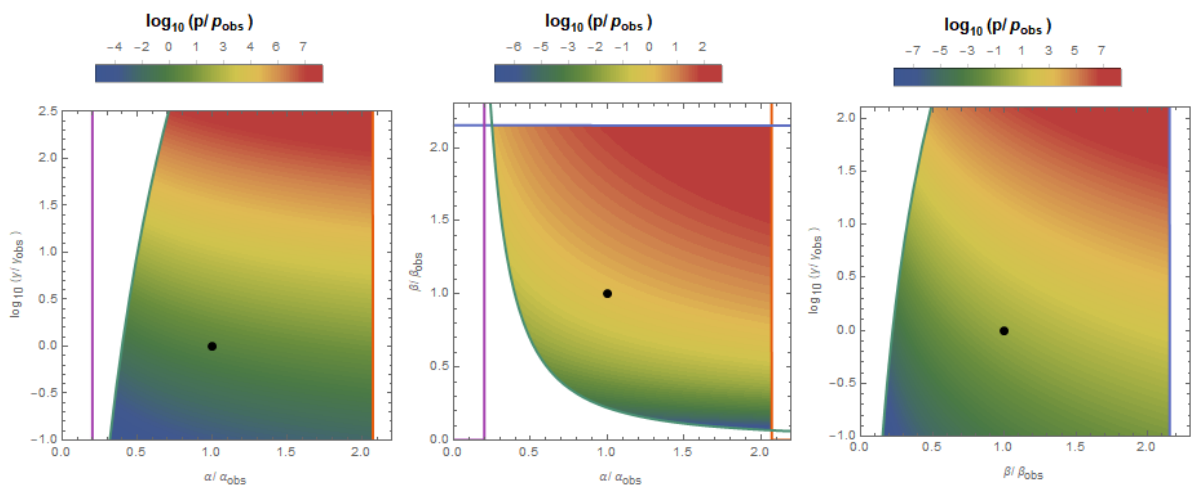

**Figure 4.** Distribution of observers if life may only exist within the temperate zone, in the $\alpha$–$\gamma$, $\alpha$–$\beta$, and $\beta$–$\gamma$ subplanes. The teal line is the boundary for which planetary disks are smaller than the temperate zone.

Let us also take this opportunity to determine what values of parameters would render disks smaller than the habitable orbit, thereby precluding temperate planets from ever forming. Based off the expressions above and in the appendix, we have

$$\frac{a_{\text{temp}}}{R_{\text{disk}}} = 4.0 \times 10^5 \, \kappa \, \lambda^{17/12} \, \frac{\gamma^{1/2}}{\alpha^5 \beta^2}. \tag{37}$$

Aside from the obvious condition that, if the mean free path of the galaxy were 100 times smaller, the protoplanetary disk would be shrunk by the same amount, this also places restrictions on the fundamental parameters. Though these introduce lower bounds for $\alpha$ and $\beta$, this is the region with very few temperate observers anyway, and so this does not change the habitability estimate appreciably, especially since these scales would have to shift by two orders of magnitude before the threshold is reached. A more stringent boundary would require that disks be larger than the orbits of the gas giants. If not, they would be precluded from forming in the first place, and, if these were essential for life on Earth, universes like this would be altogether barren.

*3.3. Planet Migration*

Traditional models of planet formation assume that little orbital migration takes place during planet formation. However, the large number of gas and ice giants found situated extremely close to their host stars [82], the presence of orbital resonances found in some exoplanet systems [83], and icy composition of planets within the snow line [84] all point to the presence of migration. Migration may strongly affect the habitability of planetary systems, as if it proceeds for too long, all inner planets will drift toward the inner edge of the disk at around 0.014 AU [85], and giant planets migrating across the temperate zone would destabilize the orbits of any existing planets. The vast diversity of systems found, as well as analytic and numeric simulations of protoplanetary disks [86], point to an exquisitely complex array of migration scenarios, which will be selectively operational depending on the characteristics of the initial system such as disk mass, viscosity, and the size and locations of the planets. Nevertheless, it is possible to perform rough estimates for when migration will be present in a given system. In this section, we do not attempt to characterize all the complex features of planet migration in universes with alternate physical constants, but rather wish to provide a useful diagnostic for when migration will be important. We will find conditions that the physical constants must satisfy so that all planets do not migrate into their host stars at an early stage of evolution, our observed values being intermediate such that this scenario only afflicts some percentage of planetary systems with unlucky characteristics. It should be noted that some amount of migration appears to have occurred even in the outer solar system [87]. However, we evidently ended up with at least one habitable location. In our particular instance, the particular positions of the giant planets ultimately halted migration, preventing the obliteration of the inner solar system [88].

Migration occurs when a planet's influence on the surrounding disk produces a net torque on the planet itself, usually driving it inward. As such, migration halts after the time $t_{\rm disk}$ when the disk has cleared out. This may be compared to the migration timescale

$$t_{\rm mig} \sim \frac{a}{\dot a} \sim \frac{L}{\Gamma} \sim \frac{\Omega\, a^2\, M}{\Gamma}, \tag{38}$$

where $M$ is the planet's mass, $L$ its angular momentum and $\Gamma$ the torque it experiences. A heuristic condition for when migration will be significant was found in [89] as $t_{\rm mig} \lesssim 10\, t_{\rm disk}$. There are various contributions to the torque, and which is dominant organizes migration into several different types, which depend on the circumstances of the case at hand. These are classified into type I, which arises when a trailing overdensity of dust behind a planet (and preceding in front) exerts a torque, and type II, in which the planet is capable of opening up a gap in the disk, resulting in a torque imbalance from the absence of material (for a recent review see [86]). The latter is more relevant for larger planets capable of significantly altering the disk structure, and the former more relevant for smaller planets, which are not. They will each be considered in turn, resulting in conditions on the fundamental constants that must be satisfied in order for a system with given characteristics to retain its initially habitable planets.

For type I migration, the timescale is derived in [90]:

$$t_{\rm I} = \frac{h^2\, M_\star^2}{\Sigma\, \Omega\, a^4\, M}, \tag{39}$$

where $h$ is the disk height, set by the sound speed. Then, when using the expressions from the Appendix A and Equation (10), and specifying to terrestrial planets situated in the temperate zone, we find

$$\frac{t_{\rm I}}{t_{\rm disk}} \propto \frac{\lambda^{43/48}\, \beta^{9/16}\, \gamma^{3/8}}{\kappa\, \alpha^{3/2}}. \tag{40}$$

If this quantity becomes too large, then Earthlike planets in all systems will migrate into their stars, and the universe would be uninhabitable in the traditional sense. Note the dependence $\lambda^{0.9}$, indicating that this type of migration is more important for low mass stars.

This estimate would also be relevant for the production mechanism for Trappist-1 type planets proposed in [91], whereby earthlike planets form behind the snow line before migrating inwards. Such an unconventional pathway is needed to explain this system [92], which has multiple Earth sized planets orbiting the temperate zone of a 0.08 solar mass star with planets near mean motion resonances [93] and possessing icy compositions [84]. This will not be undertaken here, however.

Even if we restrict our attention to parameter values where Earthlike planets do not undergo substantial migration, we may still run into trouble if Jupiter-like planets routinely barrel through their planetary systems. As this is governed by the physics of type II migration, the conditions that must be satisfied for (the majority of) these planets to stay put are somewhat different. Here, we restrict our attention to the case where a full gap is opened in the disk, and where the mass of the planet is substantially smaller than the mass contained in the disk. In this case, the timescale of this migration process is given by [86]:

$$t_{\text{II}} = \frac{2}{3}\frac{a^2}{\nu},\tag{41}$$

where $\nu$ is the viscosity. Strictly speaking, these quantities are supposed to be evaluated not at the position of the planet, but rather the place of maximum angular momentum deposition, which can be approximated as $a + 2.5 R_{\text{Hill}}$. For our purposes, however, this correction is negligible for the scaling arguments. Then, the ratio of this timescale to the disk lifetime is

$$\frac{t_{\text{II}}}{t_{\text{disk}}} \propto \frac{\lambda^{15/16}\gamma^{5/8}}{\alpha_{\text{disk}}\,\alpha^4\,\beta^{13/16}}.\tag{42}$$

Here, $\alpha_{\text{disk}}$ is the standard parameterization of disk viscosity, discussed further in the appendix. Since this is inversely proportional to disk mass, type II migration is more important for smaller disks, opposite to the type I case. Additionally, since only the scaling with $\beta$ is flipped from before, the conjunction of these two migration scenarios is capable of putting an upper bound on $\alpha$ and a lower bound on $\gamma$, if the other quantity is fixed. The thresholds for both types of migration are displayed in Figure 5. However, the absolute normalization of each of these timescales is uncertain, so we do not derive how the probabilities are altered due to these effects.

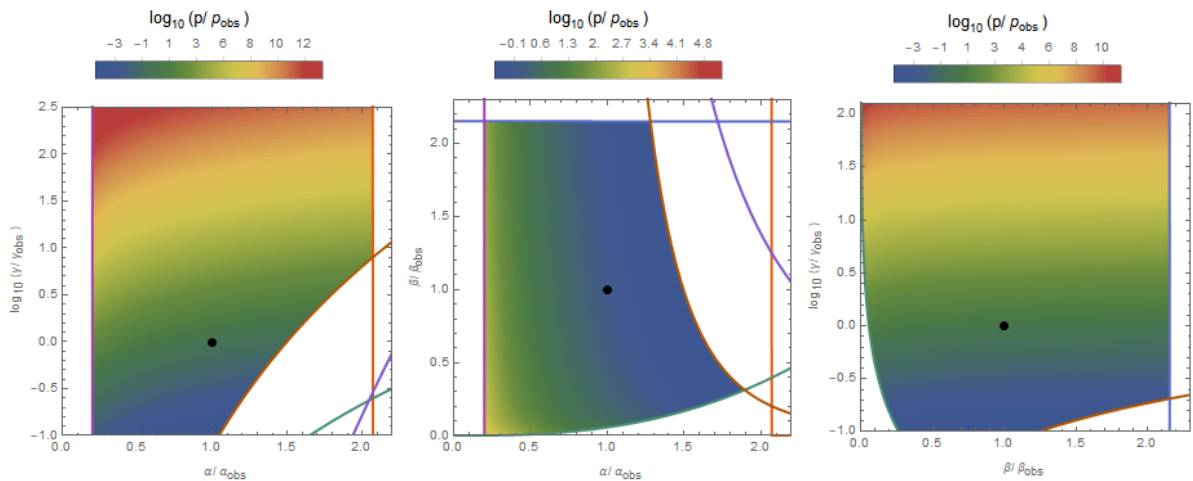

**Figure 5.** Display of when runaway migration takes hold, in the $\alpha$–$\gamma$, $\alpha$–$\beta$, and $\beta$–$\gamma$ subplanes. The teal line is for type I migration, the orange line is for type II, and the purple line is when the mass of the fastest migrating body is equal to the terrestrial mass. All curves have been normalized so the timescales in our universe are five times larger than the critical value.

**4. Discussion: Comparing 480 Hypotheses**

Having spent the last two sections detailing the physics behind a multitude of processes that may influence the habitability of a system, we now synthesize these into an estimate of the probability of observing our measured parameter values, for each combination of individual habitability hypotheses. To summarize our results so far, we have firstly included several conditions necessary for planet formation, namely that the majority of galaxies should be larger than the minimal retentive mass that massive stars should have shorter lifetimes than the star formation timescale, and that the minimum metallicity needed to form planets should be smaller than the asymptotic value. Of these three, the third was most constraining. These are also presumably not optional, as opposed to the rest of the criteria that were considered: these were the absence of hot Jupiters, the production of terrestrial planets both through giant impact and isolation formation pathways, and the fraction of planets that end up in the temperate zone. Whether these are necessary for life are still intensely debated, and so each provides a prime opportunity to determine its compatibility with the multiverse hypothesis. Because they are all independent criteria, taken in conjunction they lead to a total of $2 \times 3 \times 2 = 12$ possibilities. This does not include the 16 different choices we made in terms of planet formation, as well as the potentially continuous parameter signifying the slope of the power law for smaller planets. We display the probabilities of observing our measured values in Table 4 for the criteria mentioned in this paper. Note that, though the spread in probabilities is around 2–3 for each, all choices are well within an acceptable range to explain our observations within the multiverse context.

**Table 4.** Probabilities of different combinations of the habitability hypotheses discussed in the text. Here, yellow stands for the photosynthesis criterion, S the entropy criterion (both explained below Equation (1)), temp for the temperate zone, GI and iso for the terrestrial planet criterion with the giant impact and isolation production mechanisms, and HJ the hot Jupiter condition.

| Criteria | $\mathbb{P}(\alpha_{\text{obs}})$ | $\mathbb{P}(\beta_{\text{obs}})$ | $\mathbb{P}(\gamma_{\text{obs}})$ | $\mathbb{L}$ |
|---|---|---|---|---|
| yellow S | 0.189 | 0.437 | 0.318 | 2.57 |
| yellow temp S | 0.241 | 0.363 | 0.151 | 1.54 |
| yellow GI S | 0.229 | 0.26 | 0.409 | 6.72 |
| yellow GI temp S | 0.377 | 0.196 | 0.42 | 6.65 |
| yellow iso S | 0.234 | 0.419 | 0.436 | 5.02 |
| yellow iso temp S | 0.313 | 0.48 | 0.245 | 3.54 |
| yellow HJ S | 0.181 | 0.428 | 0.308 | 2.59 |
| yellow HJ temp S | 0.237 | 0.359 | 0.147 | 1.54 |
| yellow HJ GI S | 0.23 | 0.261 | 0.41 | 6.72 |
| yellow HJ GI temp S | 0.378 | 0.197 | 0.419 | 6.64 |
| yellow HJ iso S | 0.231 | 0.424 | 0.431 | 5.05 |
| yellow HJ iso temp S | 0.311 | 0.483 | 0.241 | 3.54 |

When combined with the 40 additional combinations from [1] (including combinations of the tidal locking criterion, which posits that only planets that are not tidally locked are habitable, the biological timescale criterion, positing that only stars that last several billion years are habitable, the convective criterion, which states that only stars which are not purely convective are habitable, the photosynthesis and yellow criteria, which state that photosynthesis is necessary for complex life, defined with an optimistic and pessimistic wavelength range, respectively, and the entropy criterion, where habitability is proportional to the total amount of entropy processed by a system), there are a total of 480 separate habitability criteria that may be considered. Of these, only 43% of them give rise to probabilities of observing all three constants we consider of greater than 1%. The full suite of criteria is displayed in Table 5 at the end of this manuscript. For brevity, we omit the convective criteria of [1] because it only ever marginally changes the numerical values, and in all instances its inclusion does not affect the viability of the combination of other hypotheses one way or the other. Of the 190 habitability criteria which give probabilities of over 10%, all make use of the entropy condition. A further 16 which do not include the entropy condition have probabilities greater than 1%—all of them benefit from an interplay between the yellow, tidal locking and biological timescale criteria, which place both upper and lower bounds on the types of allowed stars. The rest of the habitability criteria can safely be regarded as incompatible with the multiverse hypothesis.

The inclusion of multiple hypotheses leads to nonlinear effects, as the interplay between the distribution of purported observers and the anthropic boundaries alter the overall probabilities in sometimes surprising ways. That being said, none of the criteria have that drastic of an effect on the probabilities, especially when including the entropy condition.

Some criteria, namely the terrestrial and temperate conditions, introduce lower bounds to some combination of $\alpha$ and $\beta$. In fact, lower bounds on these quantities are somewhat hard to come by in the anthropics literature, though it has always been clear that they should exist, as a world with massless electrons or no electromagnetism would certainly be very different from our own. The bounds we find are stronger than those that exist in the literature.

We also introduce a new measure of our universe's fitness, which we term the luxuriance. This is defined as the expected number of observers in our universe divided by the average number of observers per universe, restricting to universes that do have observers:

$$\mathbb{L} = \frac{\mathrm{P}(\alpha_{\mathrm{obs}}, \beta_{\mathrm{obs}}, \gamma_{\mathrm{obs}}) \int d\vec{\alpha}\, \theta(\mathrm{P}(\alpha, \beta, \gamma))}{\int d\vec{\alpha}\, \mathrm{P}(\alpha, \beta, \gamma)}. \tag{43}$$

Here, the integration is over all three constants, and $\theta(x)$ is the Heaviside step function: $\theta(0) = 0$, $\theta(x) = 1$ for $x > 0$. The rationale for including this is that, for most habitability criteria, the vast majority of universes will be sterile, obfuscating comparisons between different criteria. Restricting to universes that only contain life gives a better feeling for how good our universe is at satisfying the chosen criteria. If our universe is better than typical at making life, then this quantity will be greater than 1. While this is not actually the guiding principle for evaluating whether our observations are consistent with the multiverse, it is a somewhat interesting quantity to consider. It gives some indication of how strongly observers may cluster within the multiverse- and the strong dependence of the properties we discuss on physical constants leads us to expect that they will, so that the majority of observers do find themselves in overly productive universes. The luxuriance ranges by two orders of magnitude for the different possibilities we consider, but the maximum is $\mathbb{L} = 78.4$ for the condition that includes the yellow, tidal locking, biological timescale, hot Jupiter, isolation planet production mechanism, temperate, and entropy criteria.

In addition, recall that certain choices for the physical processes involved in determining the structure of planets greatly affected some of the probabilities: the absence of a dependence on the mass of a planet with stellar mass, and too large a slope for the isolation mass, are both disfavored from the multiverse perspective. These scenarios are not otherwise excluded, but it would count as strong evidence against the multiverse if either of these were verified to be the case.

The one conclusion that should be drawn from this study is that planets are a rather generic feature, and not especially atypical for our particular values of the fundamental constants. Metal buildup is generic, massive stars burn out quickly, and disks tend to clump faster than they dissipate, so the inclusion of these criteria barely influenced the numerical values of the probabilities at all, apart from adding some mild anthropic boundaries. Furthermore, even specifying the characteristics of the planets that are formed, or considering different scenarios of their formation, did not alter the probabilities by more than a factor of few for the most part. This points to a reassuring robustness of our predictions that they are not as highly sensitive to the vagaries of incompletely known planet formation processes as one may have feared.

While it is simple to imagine a universe where planets are almost never made (and indeed before the current plethora of exoplanets was discovered many wondered if our universe was of this character), the parameter values needed to realize this possibility are quite extreme. Additionally, there are several features of planets in our universe that are tantalizingly coincidental, seeming to beckon for an anthropic explanation: the tendency to produce Earth mass planets, the similarity of the interplanetary spacing around sunlike stars and the width of the habitable zone, and the small but nonzero fraction of stars with hot Jupiters. Nevertheless, this reasoning was not borne out when incorporating these criteria into our analysis. The multiverse hypothesis is consistent with the expectation that life may only arise on temperate, terrestrial planets in systems without hot Jupiters, but it is essentially as compatible with the complete converse. The existence of suitable planetary environments thus does not seem to be the most important factor in determining which of the potential universes we find ourselves situated in. The other factors of the Drake equation which we will explore in subsequent works [94,95] will uncover many additional predictions for the requirements of life.

## 5. Conclusions

We have demonstrated that there are plenty of habitability conditions that are completely incompatible with the multiverse: what this illustrates is that, if any of the ones we have uncovered so far are shown to be the correct condition for the emergence of intelligent life, then we will be able to conclude to a very high degree of confidence (up to $5.2\sigma$) that the multiverse must be wrong. It should be stressed that there is a great deal more that these conditions omit: nothing at all is said about how habitability is affected by things like planetary eccentricity, elemental composition, water abundance, or a host of other potentially paramount aspects of a planetary system [96,97]. The de facto stance on all omissions is that they have no bearing on habitability, and it will only be through future work, including all possibly relevant aspects that a fully coherent list of predictions may be assembled. Further still, placing a priority on the relative availability of each type of universe based on reasonably generic arguments, the precise probabilities, and the conclusions that follow will tremendously benefit from a way of being able to derive this prior with absolute surety. Since only a single one of the myriad habitability criteria is ultimately true, and since we will eventually be able to determine which one that is once we have a large enough sample of life-bearing planets, this demonstrates that the multiverse is capable of generating strong experimentally testable predictions that are capable of being verified or falsified on a reasonable timescale, the hallmark of a sensible scientific theory.

**Table 5.** Probabilities of various hypotheses, including those from [1] (continued on following pages). In addition to the abbreviations from Table 4, the shorthand is: photo: photosynthesis (optimistic), yellow: photosynthesis (conservative), TL: tidal locking, bio: biological timescale, S: entropy, temp: temperate zone, GI and iso: terrestrial planet with giant impact and isolation production mechanisms, resp. and HJ: hot Jupiter.

| Criteria | $\mathbb{P}(\alpha_{\text{obs}})$ | $\mathbb{P}(\beta_{\text{obs}})$ | $\mathbb{P}(\gamma_{\text{obs}})$ | $\mathbb{L}$ |
|---|---|---|---|---|
| number of stars | 0.381 | 0.355 | $8.06 \times 10^{-7}$ | 0.000199 |
| temp | 0.175 | 0.0524 | $8.87 \times 10^{-6}$ | 0.000923 |
| GI | 0.424 | 0.281 | $1.16 \times 10^{-6}$ | 0.00021 |
| GI temp | 0.227 | 0.279 | $9.7 \times 10^{-6}$ | 0.00269 |
| iso | 0.422 | 0.493 | $8.69 \times 10^{-7}$ | 0.000239 |
| iso temp | 0.198 | 0.112 | $8.9 \times 10^{-6}$ | 0.00158 |
| HJ | 0.366 | 0.336 | $7.14 \times 10^{-7}$ | 0.000199 |
| HJ temp | 0.175 | 0.0523 | $8.87 \times 10^{-6}$ | 0.00092 |
| HJ GI | 0.425 | 0.282 | $1.15 \times 10^{-6}$ | 0.00021 |
| HJ GI temp | 0.227 | 0.279 | $9.7 \times 10^{-6}$ | 0.0027 |
| HJ iso | 0.419 | 0.499 | $8.53 \times 10^{-7}$ | 0.00024 |
| HJ iso temp | 0.198 | 0.112 | $8.9 \times 10^{-6}$ | 0.00158 |
| bio | 0.284 | 0.106 | $2.32 \times 10^{-5}$ | 0.00379 |
| bio temp | 0.145 | 0.0152 | $3.95 \times 10^{-5}$ | 0.00282 |
| bio GI | 0.211 | 0.0161 | 0.000405 | 0.0562 |
| bio GI temp | 0.282 | 0.0162 | 0.00153 | 0.324 |
| bio iso | 0.244 | 0.216 | $5.28 \times 10^{-5}$ | 0.00961 |
| bio iso temp | 0.167 | 0.0334 | $5.42 \times 10^{-5}$ | 0.00661 |
| bio HJ | 0.284 | 0.106 | $2.03 \times 10^{-5}$ | 0.00378 |
| bio HJ temp | 0.145 | 0.0152 | $3.95 \times 10^{-5}$ | 0.00281 |
| bio HJ GI | 0.211 | 0.0162 | 0.000405 | 0.0565 |
| bio HJ GI temp | 0.282 | 0.0163 | 0.00153 | 0.325 |
| bio HJ iso | 0.244 | 0.215 | $5.13 \times 10^{-5}$ | 0.00958 |
| bio HJ iso temp | 0.167 | 0.0334 | $5.41 \times 10^{-5}$ | 0.00661 |
| TL | 0.33 | 0.455 | $4.57 \times 10^{-7}$ | $3.39 \times 10^{-5}$ |
| TL temp | 0.455 | 0.257 | $1.93 \times 10^{-7}$ | $4.61 \times 10^{-5}$ |

**Table 5.** *Cont*.

| Criteria | $\mathbb{P}(\alpha_{\mathrm{obs}})$ | $\mathbb{P}(\beta_{\mathrm{obs}})$ | $\mathbb{P}(\gamma_{\mathrm{obs}})$ | $\mathbb{L}$ |
|---|---|---|---|---|
| TL GI | 0.421 | 0.283 | $6.42 \times 10^{-7}$ | $3.09 \times 10^{-5}$ |
| TL GI temp | 0.237 | 0.29 | $8.16 \times 10^{-7}$ | 0.000146 |
| TL iso | 0.398 | 0.433 | $4.6 \times 10^{-7}$ | $5.6 \times 10^{-5}$ |
| TL iso temp | 0.454 | 0.356 | $3.63 \times 10^{-7}$ | 0.000107 |
| TL HJ | 0.306 | 0.433 | $3.11 \times 10^{-7}$ | $3.15 \times 10^{-5}$ |
| TL HJ temp | 0.454 | 0.256 | $1.8 \times 10^{-7}$ | $4.52 \times 10^{-5}$ |
| TL HJ GI | 0.422 | 0.284 | $6.38 \times 10^{-7}$ | $3.08 \times 10^{-5}$ |
| TL HJ GI temp | 0.236 | 0.291 | $8.12 \times 10^{-7}$ | 0.000146 |
| TL HJ iso | 0.393 | 0.439 | $4.39 \times 10^{-7}$ | $5.53 \times 10^{-5}$ |
| TL HJ iso temp | 0.455 | 0.355 | $3.55 \times 10^{-7}$ | 0.000106 |
| TL bio | 0.011 | 0.365 | $7.19 \times 10^{-5}$ | 0.00304 |
| TL bio temp | 0.0154 | 0.276 | $4.16 \times 10^{-5}$ | 0.00788 |
| TL bio GI | 0.0562 | 0.0184 | 0.000277 | 0.0109 |
| TL bio GI temp | 0.18 | 0.023 | 0.000429 | 0.0631 |
| TL bio iso | 0.0237 | 0.485 | 0.000107 | 0.0087 |
| TL bio iso temp | 0.0403 | 0.407 | $9.85 \times 10^{-5}$ | 0.0238 |
| TL bio HJ | 0.00968 | 0.362 | $4.84 \times 10^{-5}$ | 0.00301 |
| TL bio HJ temp | 0.0149 | 0.275 | $3.89 \times 10^{-5}$ | 0.00779 |
| TL bio HJ GI | 0.0561 | 0.0185 | 0.000276 | 0.0109 |
| TL bio HJ GI temp | 0.18 | 0.0231 | 0.000428 | 0.063 |
| TL bio HJ iso | 0.0228 | 0.488 | 0.000101 | 0.00861 |
| TL bio HJ iso temp | 0.0394 | 0.406 | $9.61 \times 10^{-5}$ | 0.0235 |
| photo | 0.439 | 0.183 | $8.43 \times 10^{-7}$ | 0.000127 |
| photo S | 0.241 | 0.382 | 0.381 | 8.95 |
| photo temp | 0.403 | 0.121 | $3.11 \times 10^{-7}$ | $7.34 \times 10^{-5}$ |
| photo temp S | 0.338 | 0.292 | 0.207 | 7.07 |
| photo GI | 0.016 | 0.296 | $8.09 \times 10^{-6}$ | 0.00169 |
| photo GI S | 0.335 | 0.142 | 0.33 | 34.8 |
| photo GI temp | 0.00942 | 0.276 | $8.96 \times 10^{-6}$ | 0.00513 |
| photo GI temp S | 0.497 | 0.113 | 0.424 | 45.7 |
| photo iso | 0.456 | 0.267 | $1.75 \times 10^{-6}$ | 0.000346 |
| photo iso S | 0.305 | 0.448 | 0.494 | 13.7 |
| photo iso temp | 0.329 | 0.18 | $6.74 \times 10^{-7}$ | 0.000207 |
| photo iso temp S | 0.403 | 0.458 | 0.321 | 12.7 |
| photo HJ | 0.439 | 0.183 | $8.22 \times 10^{-7}$ | 0.000126 |
| photo HJ S | 0.237 | 0.377 | 0.376 | 9.0 |
| photo HJ temp | 0.403 | 0.121 | $3.08 \times 10^{-7}$ | $7.31 \times 10^{-5}$ |
| photo HJ temp S | 0.337 | 0.291 | 0.205 | 7.08 |
| photo HJ GI | 0.016 | 0.297 | $8.04 \times 10^{-6}$ | 0.00168 |
| photo HJ GI S | 0.336 | 0.142 | 0.331 | 35.0 |
| photo HJ GI temp | 0.00942 | 0.276 | $8.92 \times 10^{-6}$ | 0.00513 |
| photo HJ GI temp S | 0.498 | 0.113 | 0.425 | 45.9 |
| photo HJ iso | 0.456 | 0.267 | $1.72 \times 10^{-6}$ | 0.000342 |
| photo HJ iso S | 0.302 | 0.452 | 0.497 | 13.8 |
| photo HJ iso temp | 0.329 | 0.18 | $6.68 \times 10^{-7}$ | 0.000206 |
| photo HJ iso temp S | 0.402 | 0.455 | 0.319 | 12.8 |
| photo bio | 0.0631 | 0.103 | $1.75 \times 10^{-5}$ | 0.00197 |
| photo bio S | 0.18 | 0.409 | 0.43 | 7.89 |
| photo bio temp | 0.123 | 0.0849 | $5.35 \times 10^{-6}$ | 0.000987 |
| photo bio temp S | 0.222 | 0.335 | 0.262 | 6.98 |
| photo bio GI | 0.353 | 0.0162 | 0.000924 | 0.166 |
| photo bio GI S | 0.316 | 0.131 | 0.308 | 30.9 |
| photo bio GI temp | 0.154 | 0.0141 | 0.000688 | 0.34 |
| photo bio GI temp S | 0.484 | 0.1 | 0.408 | 40.3 |
| photo bio iso | 0.104 | 0.188 | $5.14 \times 10^{-5}$ | 0.00757 |
| photo bio iso S | 0.262 | 0.42 | 0.448 | 11.6 |
| photo bio iso temp | 0.185 | 0.152 | $1.51 \times 10^{-5}$ | 0.00365 |
| photo bio iso temp S | 0.322 | 0.489 | 0.382 | 11.8 |

**Table 5.** *Cont.*

| Criteria | $\mathbb{P}(\alpha_{\text{obs}})$ | $\mathbb{P}(\beta_{\text{obs}})$ | $\mathbb{P}(\gamma_{\text{obs}})$ | $\mathbb{L}$ |
|---|---|---|---|---|
| photo bio HJ | 0.0631 | 0.103 | $1.71 \times 10^{-5}$ | 0.00195 |
| photo bio HJ S | 0.175 | 0.404 | 0.426 | 7.94 |
| photo bio HJ temp | 0.123 | 0.0848 | $5.31 \times 10^{-6}$ | 0.000983 |
| photo bio HJ temp S | 0.22 | 0.333 | 0.26 | 6.99 |
| photo bio HJ GI | 0.353 | 0.0162 | 0.000921 | 0.167 |
| photo bio HJ GI S | 0.317 | 0.132 | 0.309 | 31.0 |
| photo bio HJ GI temp | 0.154 | 0.0141 | 0.000687 | 0.34 |
| photo bio HJ GI temp S | 0.485 | 0.101 | 0.409 | 40.5 |
| photo bio HJ iso | 0.104 | 0.188 | $5.06 \times 10^{-5}$ | 0.0075 |
| photo bio HJ iso S | 0.259 | 0.424 | 0.451 | 11.7 |
| photo bio HJ iso temp | 0.185 | 0.151 | $1.5 \times 10^{-5}$ | 0.00363 |
| photo bio HJ iso temp S | 0.32 | 0.492 | 0.379 | 11.8 |
| photo TL | 0.478 | 0.232 | $7.36 \times 10^{-7}$ | 0.000102 |
| photo TL S | 0.376 | 0.423 | 0.382 | 10.8 |
| photo TL temp | 0.412 | 0.197 | $2.61 \times 10^{-7}$ | $6.01 \times 10^{-5}$ |
| photo TL temp S | 0.453 | 0.288 | 0.283 | 10.8 |
| photo TL GI | 0.016 | 0.312 | $4.34 \times 10^{-6}$ | 0.000271 |
| photo TL GI S | 0.312 | 0.431 | 0.447 | 13.9 |
| photo TL GI temp | 0.0096 | 0.306 | $1.93 \times 10^{-6}$ | 0.000332 |
| photo TL GI temp S | 0.485 | 0.317 | 0.345 | 16.7 |
| photo TL iso | 0.41 | 0.308 | $1.46 \times 10^{-6}$ | 0.000264 |
| photo TL iso S | 0.384 | 0.46 | 0.435 | 15.9 |
| photo TL iso temp | 0.323 | 0.26 | $5.48 \times 10^{-7}$ | 0.000154 |
| photo TL iso temp S | 0.447 | 0.331 | 0.365 | 15.5 |
| photo TL HJ | 0.478 | 0.232 | $6.94 \times 10^{-7}$ | $9.94 \times 10^{-5}$ |
| photo TL HJ S | 0.36 | 0.394 | 0.355 | 11.2 |
| photo TL HJ temp | 0.412 | 0.197 | $2.51 \times 10^{-7}$ | $5.92 \times 10^{-5}$ |
| photo TL HJ temp S | 0.458 | 0.276 | 0.272 | 10.9 |
| photo TL HJ GI | 0.016 | 0.312 | $4.31 \times 10^{-6}$ | 0.000269 |
| photo TL HJ GI S | 0.312 | 0.43 | 0.446 | 13.9 |
| photo TL HJ GI temp | 0.00959 | 0.306 | $1.92 \times 10^{-6}$ | 0.00033 |
| photo TL HJ GI temp S | 0.484 | 0.316 | 0.344 | 16.7 |
| photo TL HJ iso | 0.41 | 0.307 | $1.42 \times 10^{-6}$ | 0.000258 |
| photo TL HJ iso S | 0.382 | 0.452 | 0.43 | 16.0 |
| photo TL HJ iso temp | 0.323 | 0.26 | $5.37 \times 10^{-7}$ | 0.000151 |
| photo TL HJ iso temp S | 0.448 | 0.325 | 0.362 | 15.5 |
| photo TL bio | 0.0354 | 0.29 | 0.000149 | 0.017 |
| photo TL bio S | 0.252 | 0.495 | 0.423 | 14.9 |
| photo TL bio temp | 0.0406 | 0.291 | 0.000116 | 0.0252 |
| photo TL bio temp S | 0.33 | 0.376 | 0.413 | 20.9 |
| photo TL bio GI | 0.317 | 0.0329 | 0.00121 | 0.0717 |
| photo TL bio GI S | 0.104 | 0.413 | 0.385 | 17.9 |
| photo TL bio GI temp | 0.371 | 0.0357 | 0.00086 | 0.14 |
| photo TL bio GI temp S | 0.196 | 0.283 | 0.397 | 27.3 |
| photo TL bio iso | 0.087 | 0.43 | 0.000325 | 0.0495 |
| photo TL bio iso S | 0.249 | 0.5 | 0.393 | 20.4 |
| photo TL bio iso temp | 0.0982 | 0.432 | 0.00026 | 0.07 |
| photo TL bio iso temp S | 0.353 | 0.383 | 0.366 | 25.1 |
| photo TL bio HJ | 0.0347 | 0.289 | 0.000141 | 0.0167 |
| photo TL bio HJ S | 0.218 | 0.456 | 0.449 | 16.0 |
| photo TL bio HJ temp | 0.0399 | 0.29 | 0.000112 | 0.025 |
| photo TL bio HJ temp S | 0.311 | 0.354 | 0.424 | 21.6 |
| photo TL bio HJ GI | 0.316 | 0.033 | 0.00121 | 0.0715 |
| photo TL bio HJ GI S | 0.103 | 0.411 | 0.385 | 17.9 |
| photo TL bio HJ GI temp | 0.37 | 0.0359 | 0.000859 | 0.14 |
| photo TL bio HJ GI temp S | 0.196 | 0.282 | 0.398 | 27.2 |
| photo TL bio HJ iso | 0.0854 | 0.429 | 0.000318 | 0.0488 |

**Table 5.** *Cont.*

| Criteria | $\mathbb{P}(\alpha_{\text{obs}})$ | $\mathbb{P}(\beta_{\text{obs}})$ | $\mathbb{P}(\gamma_{\text{obs}})$ | $\mathbb{L}$ |
| --- | --- | --- | --- | --- |
| photo TL bio HJ iso S | 0.243 | 0.49 | 0.397 | 20.7 |
| photo TL bio HJ iso temp | 0.0966 | 0.43 | 0.000255 | 0.0693 |
| photo TL bio HJ iso temp S | 0.348 | 0.375 | 0.367 | 25.3 |
| yellow | 0.486 | 0.162 | $8.78 \times 10^{-7}$ | $9.64 \times 10^{-5}$ |
| yellow temp | 0.492 | 0.161 | $2.31 \times 10^{-7}$ | $2.61 \times 10^{-5}$ |
| yellow GI | 0.00555 | 0.292 | $6.88 \times 10^{-5}$ | 0.0127 |
| yellow GI temp | 0.00329 | 0.272 | $2.81 \times 10^{-5}$ | 0.00588 |
| yellow iso | 0.391 | 0.219 | $2.03 \times 10^{-6}$ | 0.000336 |
| yellow iso temp | 0.387 | 0.218 | $5.35 \times 10^{-7}$ | $9.14 \times 10^{-5}$ |
| yellow HJ | 0.486 | 0.162 | $8.64 \times 10^{-7}$ | $9.55 \times 10^{-5}$ |
| yellow HJ temp | 0.492 | 0.161 | $2.27 \times 10^{-7}$ | $2.58 \times 10^{-5}$ |
| yellow HJ GI | 0.00554 | 0.292 | $6.83 \times 10^{-5}$ | 0.0126 |
| yellow HJ GI temp | 0.00329 | 0.272 | $2.79 \times 10^{-5}$ | 0.00584 |
| yellow HJ iso | 0.391 | 0.219 | $1.99 \times 10^{-6}$ | 0.000333 |
| yellow HJ iso temp | 0.387 | 0.218 | $5.26 \times 10^{-7}$ | $9.06 \times 10^{-5}$ |
| yellow bio | 0.0351 | 0.102 | $1.72 \times 10^{-5}$ | 0.00178 |
| yellow bio S | 0.123 | 0.47 | 0.38 | 2.88 |
| yellow bio temp | 0.0324 | 0.0912 | $5.21 \times 10^{-6}$ | 0.000555 |
| yellow bio temp S | 0.126 | 0.406 | 0.215 | 2.05 |
| yellow bio GI | 0.0503 | 0.0133 | 0.000736 | 0.13 |
| yellow bio GI S | 0.177 | 0.25 | 0.367 | 6.9 |
| yellow bio GI temp | 0.0316 | 0.00818 | 0.000315 | 0.0632 |
| yellow bio GI temp S | 0.299 | 0.177 | 0.474 | 7.21 |
| yellow bio iso | 0.081 | 0.187 | $5.23 \times 10^{-5}$ | 0.00821 |
| yellow bio iso S | 0.174 | 0.386 | 0.498 | 5.37 |
| yellow bio iso temp | 0.0741 | 0.168 | $1.62 \times 10^{-5}$ | 0.00262 |
| yellow bio iso temp S | 0.206 | 0.424 | 0.324 | 4.38 |
| yellow bio HJ | 0.0351 | 0.102 | $1.69 \times 10^{-5}$ | 0.00176 |
| yellow bio HJ S | 0.112 | 0.46 | 0.37 | 2.91 |
| yellow bio HJ temp | 0.0324 | 0.0912 | $5.14 \times 10^{-6}$ | 0.00055 |
| yellow bio HJ temp S | 0.12 | 0.4 | 0.209 | 2.05 |
| yellow bio HJ GI | 0.0503 | 0.0134 | 0.000732 | 0.13 |
| yellow bio HJ GI S | 0.177 | 0.251 | 0.368 | 6.9 |
| yellow bio HJ GI temp | 0.0316 | 0.00819 | 0.000313 | 0.0629 |
| yellow bio HJ GI temp S | 0.299 | 0.177 | 0.473 | 7.21 |
| yellow bio HJ iso | 0.0809 | 0.187 | $5.13 \times 10^{-5}$ | 0.00814 |
| yellow bio HJ iso S | 0.169 | 0.392 | 0.493 | 5.4 |
| yellow bio HJ iso temp | 0.074 | 0.168 | $1.59 \times 10^{-5}$ | 0.0026 |
| yellow bio HJ iso temp S | 0.202 | 0.428 | 0.319 | 4.39 |
| yellow TL | 0.0303 | 0.0308 | $1.63 \times 10^{-6}$ | 0.000763 |
| yellow TL S | 0.457 | 0.377 | 0.431 | 32.6 |
| yellow TL temp | 0.0229 | 0.0251 | $4.55 \times 10^{-7}$ | 0.000253 |
| yellow TL temp S | 0.328 | 0.281 | 0.242 | 26.5 |
| yellow TL GI | 0.00345 | 0.38 | $4.84 \times 10^{-5}$ | 0.0256 |
| yellow TL GI S | 0.35 | 0.299 | 0.496 | 43.7 |
| yellow TL GI temp | 0.00196 | 0.372 | $2.04 \times 10^{-5}$ | 0.0129 |
| yellow TL GI temp S | 0.481 | 0.311 | 0.326 | 44.9 |
| yellow TL iso | 0.0249 | 0.0412 | $3.78 \times 10^{-6}$ | 0.00193 |
| yellow TL iso S | 0.488 | 0.424 | 0.469 | 44.7 |
| yellow TL iso temp | 0.0186 | 0.0335 | $1.06 \times 10^{-6}$ | 0.000636 |
| yellow TL iso temp S | 0.353 | 0.353 | 0.294 | 39.2 |
| yellow TL HJ | 0.0303 | 0.0308 | $1.58 \times 10^{-6}$ | 0.000756 |
| yellow TL HJ S | 0.474 | 0.351 | 0.407 | 33.8 |
| yellow TL HJ temp | 0.0229 | 0.0251 | $4.43 \times 10^{-7}$ | 0.000251 |
| yellow TL HJ temp S | 0.334 | 0.266 | 0.227 | 26.9 |

**Table 5.** *Cont.*

| Criteria | $\mathbb{P}(\alpha_{\text{obs}})$ | $\mathbb{P}(\beta_{\text{obs}})$ | $\mathbb{P}(\gamma_{\text{obs}})$ | $\mathbb{L}$ |
| --- | --- | --- | --- | --- |
| yellow TL HJ GI | 0.00345 | 0.38 | $4.82 \times 10^{-5}$ | 0.0254 |
| yellow TL HJ GI S | 0.35 | 0.298 | 0.495 | 43.6 |
| yellow TL HJ GI temp | 0.00196 | 0.372 | $2.03 \times 10^{-5}$ | 0.0128 |
| yellow TL HJ GI temp S | 0.481 | 0.31 | 0.325 | 44.7 |
| yellow TL HJ iso | 0.0249 | 0.0412 | $3.69 \times 10^{-6}$ | 0.00191 |
| yellow TL HJ iso S | 0.493 | 0.416 | 0.463 | 45.1 |
| yellow TL HJ iso temp | 0.0186 | 0.0335 | $1.04 \times 10^{-6}$ | 0.00063 |
| yellow TL HJ iso temp S | 0.355 | 0.346 | 0.289 | 39.4 |
| yellow TL bio | 0.324 | 0.335 | 0.0114 | 5.54 |
| yellow TL bio S | 0.373 | 0.496 | 0.323 | 51.3 |
| yellow TL bio temp | 0.332 | 0.335 | 0.00786 | 4.56 |
| yellow TL bio temp S | 0.399 | 0.474 | 0.425 | 63.1 |
| yellow TL bio GI | 0.472 | 0.446 | 0.0126 | 6.64 |
| yellow TL bio GI S | 0.126 | 0.278 | 0.323 | 59.4 |
| yellow TL bio GI temp | 0.452 | 0.444 | 0.0093 | 5.88 |
| yellow TL bio GI temp S | 0.183 | 0.283 | 0.433 | 77.6 |
| yellow TL bio iso | 0.435 | 0.446 | 0.0158 | 8.4 |
| yellow TL bio iso S | 0.349 | 0.492 | 0.326 | 64.5 |
| yellow TL bio iso temp | 0.444 | 0.445 | 0.0108 | 6.82 |
| yellow TL bio iso temp S | 0.404 | 0.499 | 0.423 | 77.2 |
| yellow TL bio HJ | 0.322 | 0.333 | 0.0111 | 5.5 |
| yellow TL bio HJ S | 0.332 | 0.461 | 0.343 | 54.5 |
| yellow TL bio HJ temp | 0.331 | 0.333 | 0.00769 | 4.53 |
| yellow TL bio HJ temp S | 0.369 | 0.447 | 0.444 | 66.0 |
| yellow TL bio HJ GI | 0.472 | 0.446 | 0.0126 | 6.61 |
| yellow TL bio HJ GI S | 0.126 | 0.277 | 0.323 | 59.3 |
| yellow TL bio HJ GI temp | 0.453 | 0.445 | 0.00928 | 5.85 |
| yellow TL bio HJ GI temp S | 0.183 | 0.281 | 0.434 | 77.4 |
| yellow TL bio HJ iso | 0.433 | 0.444 | 0.0155 | 8.36 |
| yellow TL bio HJ iso S | 0.339 | 0.499 | 0.33 | 65.5 |
| yellow TL bio HJ iso temp | 0.442 | 0.443 | 0.0106 | 6.8 |
| yellow TL bio HJ iso temp S | 0.395 | 0.492 | 0.428 | 78.4 |

**Funding:** This research received no external funding.

**Acknowledgments:** I would like to thank Cullen Blake, Diana Dragomir, Scott Kenyon, Jabran Zahid, and Li Zeng for useful discussions.

**Conflicts of Interest:** The author declares no conflict of interest.

## Appendix A. Planetary Parameters

In this appendix, we collect results on how various quantities relevant to our estimates in this work depend on orbital parameters of the stellar system, as well as fundamental quantities. To begin, we display the typical molecular binding energy:

$$T_{\text{mol}} = \sqrt{\frac{\alpha}{m_{\text{mol}} r_{\text{mol}}^3}} = 0.037 \, \frac{\alpha^2 \, m_e^{3/2}}{m_p^{1/2}}. \tag{A1}$$

This also defines the temperature required for liquid water.

**Planets:** The mass of a terrestrial planet, based on the criteria that carbon dioxide but not helium is gravitationally bound to the surface at these temperatures, is

$$M_{\text{terr}} = 9.2 - 202.2 \, \frac{\alpha^{3/2} \, m_e^{3/4} \, M_{pl}^3}{m_p^{11/4}}. \tag{A2}$$

Next, we display the range of habitable orbits from the host star, using estimates from [1,98]:

$$a_{\text{temp}} = 7.2 - 12.3 \, \lambda^{7/4} \, \frac{m_p^{1/2} \, M_{pl}^{1/2}}{\alpha^5 \, m_e^2}, \tag{A3}$$

which is based off of the temperature of a perfect blackbody located at that position,

$$T(a) = T_\star \sqrt{\frac{R_\star}{2a}}. \tag{A4}$$

The speed of a circularly orbiting planet at a given location is

$$v = \sqrt{\frac{GM_\star}{a}} = \Omega \, a \tag{A5}$$

and $\Omega$ is the angular frequency. This also defines the orbital period as $t_{\text{Kepler}} = 2\pi/\Omega$.

The region of influence of a planet of mass $M$ and orbiting a star at semimajor axis $a$ is known as the Hill sphere, and has the radius

$$R_{\text{Hill}} = \left(\frac{M}{3 \, M_\star}\right)^{1/3} a. \tag{A6}$$

**Galaxies:** the average galactic density is set by the galaxy at time of virialization, with a subsequent era of contraction due to equilibration [11,99]:

$$\rho_{\text{gal}} = 5.6 \times 10^7 \, \kappa^3 \, m_p^4 \tag{A7}$$

corresponding to $10^{-23}$ g/cm$^3$. Here, $\kappa = Q(\eta \omega)^{4/3} = 10^{-16}$ is a composite of the primordial amplitude of perturbations $Q = 1.8 \times 10^{-5}$, the baryon to photon ratio $\eta = 6 \times 10^{-10}$, and the total matter to baryon ratio $\omega = 6.4$. This defines the typical freefall timescale as

$$t_{\text{ff}} = \frac{1}{\sqrt{G \rho_{\text{gal}}}} = 6.7 \times 10^{-4} \frac{M_{pl}}{\kappa^{3/2} \, m_p^2}, \tag{A8}$$

which equates to $3 \times 10^7$ yr. In addition, the typical temperature of the interstellar gas is relevant, which is set by the threshold for H$_2$ cooling, at roughly $10^4$ K:

$$T_{\text{H}_2} = 0.032 \, \alpha^2 \, m_e. \tag{A9}$$

This also sets the velocity dispersion in the galaxy as $v \sim \sqrt{5 T_{\text{H}_2}/m_p} \sim 20$ km/s.

**Disks:** We take the disk mass to be set by

$$M_{\text{disk}} = 0.01 \, M_\star. \tag{A10}$$

The size of the disk is set by the conservation of angular momentum of the initial collapsing cloud. If it has a typical angular momentum of $L \sim 0.01 Mvr$, with $v$ given by the dispersion velocity of molecular clouds, we find

$$r_{\text{disk}} \sim \left(\frac{M_\star}{\rho_{\text{gal}}}\right)^{1/3} = 3.6 \times 10^{-6} \frac{\lambda^{1/3} M_{pl}}{\kappa \, m_p^2} \tag{A11}$$

set to be 100 AU for sunlike stars.

For infalling dust, the accretion rate $\dot{M}$ is just given in terms of the typical free fall timescale as [100]

$$\dot{M} = \frac{c_s^3}{G_N} = 7.2 \times 10^{-4} \frac{\lambda^2 \alpha^3 m_e^{9/4} M_{pl}^2}{m_p^{9/4}}. \tag{A12}$$

In evaluating this quantity, the molecular energy has been used, reflecting that star formation exclusively forms within molecular clouds, where increased levels of cooling facilitate collapse. A disk whose dominant form of heating is given by accretion will have a temperature set by [55]

$$T_{\text{accr}}^4 = \frac{3}{8\pi} \frac{G M_\star \dot{M}}{a^3}. \tag{A13}$$

The speed of sound in the disk is given by $c_s \sim \sqrt{T(a)/m_p}$. This sets the height of the disk as the typical sound dispersion during one oscillation period. Using the accretion temperature,

$$h \sim \frac{c_s}{\Omega} = .34 \frac{\alpha^{3/8} m_e^{9/32} a^{9/8}}{\lambda^{1/8} m_p^{1/32} M_{pl}^{1/8}}. \tag{A14}$$

This is normalized to 0.08 AU for earthlike orbits around sunlike stars. The viscosity of the disk is usually parameterized as

$$\nu = \frac{\alpha_{\text{disk}} c_s^2}{\Omega}. \tag{A15}$$

Typical values for the coefficient of proportionality are $\alpha_{\text{disk}} \sim 10^{-3} - 10^{-2}$. In equilibrium, the surface density of the disk can be found to be given by

$$\Sigma = \frac{\dot{M}}{3\pi \nu}. \tag{A16}$$

The mass loss is usually taken to be independent of position [60], so that the surface density profile is dictated by the temperature. Early models took $\Sigma \propto a^{-3/2}$, but this is now considered unlikely for equilibrium disks. The now standard dependence is $\Sigma \propto 1/a$ [51], which is both observed in simulations and understood from a theoretical perspective [54]. We will take our profile to be

$$\Sigma(a) = \frac{M_{\text{disk}}}{2 r_{\text{disk}} a} \theta(r_{\text{disk}} - a). \tag{A17}$$

The snow line is the distance beyond which the disk is cool enough for water to condense into a solid phase, which is found by setting the temperature equal to the vibrational energy of water molecules,

$$a_{\text{snow}} = 30.4 \frac{\lambda M_{pl}}{\alpha^{5/3} m_e^{5/4} m_p^{3/4}}. \tag{A18}$$

The size a planetesimal attains is given by the isolation mass, $M_{\text{iso}} \sim 10 \times 2\pi \Sigma a R_{\text{Hill}}$, [60]. The rationale behind this quantity is that, once a planet is above this mass, it will have already depleted all the material within its Hill radius, and so it will no longer have a supply to continue its growth. Note that the Hill radius is itself a function of the planet mass, so that solving this equation for the mass yields the expression:

$$M_{\text{iso}} \sim \frac{(2\pi \Sigma a^2)^{3/2}}{M_\star^{1/2}} = 1.2 \times 10^6 \kappa^{3/2} \lambda^{1/2} m_p M_{pl}^{3/2} a^{3/2}. \tag{A19}$$

The coefficient is set by noting that the isolation mass is about Mars sized for the solar system.